\documentclass[useAMS,usenatbib]{mn2e}

\usepackage{graphicx}
\usepackage{aas_macros}
\usepackage{amsfonts,amssymb}
\usepackage{color}

\newcommand{\lya}{Ly$\alpha$}
\newcommand{\lyb}{Ly$\beta$}

\newcommand{\hi}{H~{\sc i}}
\newcommand{\oi}{O~{\sc i}}
\newcommand{\cii}{C~{\sc ii}}

\newcommand{\civ}{C~{\sc iv}}
\newcommand{\siii}{Si~{\sc ii}}

\newcommand{\siiv}{Si~{\sc iv}}
\newcommand{\nv}{N~{\sc v}}

\newcommand{\mgii}{Mg~{\sc ii}}

\newcommand{\ovi}{O~{\sc vi}}

\newcommand{\mhi}{{\rm H \; \mbox{\tiny I}}}

\newcommand{\kms}{km~s$^{-1}$}

\newcommand{\taueff}{$\tau_{\rm eff}$}
\newcommand{\taueffz}{$\tau_{\rm eff}(z)$}

\newcommand{\meanfluxz}{$F (z)$}
\newcommand{\fg}{Faucher-Gigu\`ere et al.}
\newcommand{\bernardi}{Bernardi et al.}
\newcommand{\da}{Dall'Aglio et al.}
\newcommand{\paris}{P\^aris et al.}

\title[Mean Transmitted \lya\ Flux over $2 < z < 5$]{A Refined Measurement of the Mean Transmitted Flux in the \lya\ Forest over $2 < z < 5$ Using Composite Quasar Spectra}

\author[Becker et al.]
   {George D.~Becker$^1$\thanks{gdb@ast.cam.ac.uk},
   Paul C.~Hewett$^1$, G\'abor Worseck$^2$ and J.~Xavier~Prochaska$^2$ \\
   $^1$Kavli Institute for Cosmology and Institute of Astronomy, Madingley Rd, Cambridge, CB3 0HA, UK \\
   $^2$Department of Astronomy and Astrophysics, UCO/Lick Observatory, University of California, 1156 High Street, Santa Cruz, \\ CA 95064, USA}
   \date{Draft version \today}
   
\begin{document}

\label{firstpage}

\maketitle

\begin{abstract}  

We present new measurements of the mean transmitted flux in the \lya\ forest over $2 < z < 5$ made using 6065 quasar spectra from the Sloan Digital Sky Survey DR7.  We exploit the general lack of evolution in the mean quasar continuum to avoid the bias introduced by continuum fitting over the \lya\ forest at high redshifts, which has been the primary systematic uncertainty in previous measurements of the mean \lya\ transmission.  The individual spectra are first combined into twenty-six composites with mean redshifts spanning $2.25 \le z_{\rm comp} \le 5.08$.  The flux ratios of separate composites at the same rest wavelength are then used, without continuum fitting, to infer the mean transmitted flux, $F(z)$, as a fraction of its value at $z \sim 2$.  Absolute values for $F(z)$ are found by scaling our relative values to measurements made from high-resolution data by \citet{fg2008a} at $z \le 2.5$, where continuum uncertainties are minimal.  We find that $F(z)$ evolves smoothly with redshift, with no evidence of a previously reported feature at $z \simeq 3.2$.  This trend is consistent with a gradual evolution of the ionization and thermal state of the intergalactic medium over $2 < z < 5$.  Our results generally agree with the most careful measurements to date made from high-resolution data, but offer much greater precision and extend to higher redshifts.  This work also improves upon previous efforts using SDSS spectra by significantly reducing the level of systematic error.  

\end{abstract}


\section{Introduction}

The pattern of absorption lines imprinted on the spectra of distant objects by neutral hydrogen in the intergalactic medium (IGM), known as the ``\lya\ forest'', is one of the most fundamental probes of cosmic structure.  The characteristics of this absorption reflect the density distribution, ionization state, and temperature of the intergalactic gas.  The gas, in turn, closely traces the underlying distribution of dark matter, albeit with significant deviations arising from hydrodynamics coupled to the radiative and mechanical feedback from galaxies and AGN.  Consequently, the \lya\ forest allows us to probe the baryon physics connecting the evolution of large-scale structure to the highly non-linear processes driving galaxy formation.

A key requirement for utilizing the \lya\ forest to study either cosmology or baryon physics is to accurately establish the basic parameters describing the flux distribution.  The most basic of these is the mean transmitted flux, $F = \langle e^{-\tau} \rangle$, and its evolution with redshift.  The mean flux, which is often expressed in terms of an effective optical depth, $\tau_{\rm eff}(z) =  -\ln{F(z)}$, directly constrains, for example, the intensity of the metagalactic ionizing background \citep[e.g.,][]{rauch1997,mcdonald2001a,bolton2005}.  It also influences statistics such as the flux probability distribution function and power spectrum, among others, which are used to constrain quantities ranging from the temperature of the IGM \citep[e.g.,][]{schaye2000,zaldarriaga2001,lidz2010,becker2011a} to the free-streaming length of dark matter particles \citep[e.g.,][]{viel2008}.  \lya\ forest studies increasingly rely on comparing real data to artificial absorption spectra drawn from simulations, which must be calibrated to the correct mean flux.  With the increasing availability of high-quality quasar and gamma ray burst spectra taken with large telescopes, and the large number of moderate-resolution quasar spectra gathered with surveys such as the Sloan Digital Sky Survey \citep[SDSS;][]{york2000} and the Baryon Oscillation Spectroscopic Survey \citep[BOSS;][]{dawson2012}, a precise measurement of $F(z)$ is therefore critical for accurately extracting the maximum amount of information from the forest. 

The main challenge in measuring \meanfluxz\ is to overcome cosmic variance while accurately estimating the unabsorbed continuum of the background objects, which have so far typically been quasars.  Studies using high-resolution (generally $R \sim 40,000$) spectra have the advantage that the peaks of the \lya\ forest transmission should approach the continuum \citep[e.g.,][]{rauch1997,mcdonald2000,schaye2003,songaila2004,kirkman2005,kim2007,becker2007,fg2008a,dallaglio2008,becker2011a}.  Even at moderate redshifts ($z \sim2-3$), however, voids in the IGM will have a non-negligible optical depth.  The continuum will therefore lie somewhat above the tops of the transmission peaks, a bias that increases towards higher redshifts.  Alternatively, studies using large sets of moderate-resolution spectra have either attempted to simultaneously solve for \meanfluxz\ and the quasar continua in a strictly statistical sense \citep{bernardi2003}, or have fit continua on an individual basis using principle component analysis \citep[PCA;][]{mcdonald2005,paris2011} or with bias corrections estimated from model spectra \citep{dallaglio2009}.  At the highest redshifts ($z > 5$), \meanfluxz\ measurements have generally relied on extrapolating a power-law continuum from redward of the \lya\ emission line \citep{songaila2004,fan2006b,becker2007}.

The most accurate \meanfluxz\ measurement to date has arguably been made by \citet{fg2008a}.  These authors used a set of 86 high-resolution quasar spectra and made statistical corrections to the continua as a function of redshift based on artificial spectra drawn from hydrodynamic simulations.  Notably, they identified a possible sharp ($\Delta z \lesssim 0.5$) dip in the evolution of the mean opacity near $z \simeq 3.2$.  A similar feature was found by \citet{bernardi2003} using a very different data set and measurement technique.  The optical depth of the optically thin IGM to \lya\ will scale with the \hi\ neutral fraction, which in turn depends on the gas temperature, $T$, and the \hi\ ionization rate, $\Gamma$, as $f_{\mhi} \propto T^{-0.7}\Gamma^{-1}$, where the temperature dependence is for case A recombination.  \bernardi\ noted that a  decrease in \taueffz\ at $z \simeq 3.2$ could indicate a temporary increase in the IGM gas temperature accompanying the end of helium reionization \citep[see also][]{theuns2002c,fg2008a}.  However, the reality of this feature remains unclear.  Although a similar dip was found at modest significance by \citet{dallaglio2008} in a sample of 40 high-resolution spectra, neither \citet{mcdonald2005} nor \citet{dallaglio2009} detected such a feature using samples of moderate-resolution SDSS spectra larger than that of  \citet{bernardi2003}.  Indeed, \citet{fg2008a} emphasized that the significance of their feature may have been exaggerated by the assumption of an underlying power law evolution in \taueff.  The association of such a sharp feature with helium reionization has also been challenged on theoretical grounds.  \citet{bolton2009b} argued that even if the IGM is photo-heated during the course of a very rapid ($\Delta z \simeq 0.2$) helium reionization, the expected signature would be a sudden decrease in \taueff\ followed by a gradual recovery driven by adiabatic cooling in the voids.  Moreover, recent temperature measurements indicate that heating due to helium reionization was an extended process with $\Delta z \gtrsim 1$ \citep{becker2011a}, and so it is unclear why such a sharp feature in the opacity evolution should occur.

In this paper we present a new and highly precise measurement of the mean transmitted \lya\ flux over $2 < z < 5$ using a sample of 6065 moderate-resolution quasar spectra drawn from the seventh data release of the Sloan Digital Sky Survey (SDSS DR7).  The main innovation of this work is the introduction of a new method  that uses composite spectra to measure the overall shape the mean flux evolution to high accuracy without fitting continua in the \lya\ forest.  These results are then normalized to measurements made from high-resolution data at $z \sim 2$, where continuum errors are minimal.  In common with all strategies that attempt to estimate quasar continua over the \lya\ forest based on redder, unabsorbed regions of the spectra (e.g., PCA analysis), it is essentially impossible to be completely certain that redshift-dependent variations in quasar spectral energy distributions (SEDs) are not affecting the derived $F(z)$ measurements.  We present tests, however, which demonstrate that such systematics are unlikely to be affecting our results at a level greater than the error bars. Our approach builds on the successful use of composite quasar spectra to statistically constrain other properties of the IGM, for example the mean free path of ionizing photons \citep{prochaska2009b}.  We also take advantage of the improved statistical accuracy afforded by DR7 to increase substantially the number of quasars analyzed compared to previous works \citep[][although \paris\ also used DR7 data]{bernardi2003,mcdonald2005,dallaglio2009,paris2011}.  In Section~\ref{sec:data_method} we describe our technique, which uses composite spectra to compute \meanfluxz\ relative to that at $z \sim 2$.  Our results are presented in Section~\ref{sec:results}.  In Section~\ref{sec:comparison}, we compare our measurements to results from the literature.  Finally, we summarize in Section~\ref{sec:summary}.

\section{Data \& Method}\label{sec:data_method}

The normalized transmitted flux in a single spectrum is given by the observed flux divided by the unabsorbed continuum, $F^i = f_{\rm obs}^i / f_{\rm cont}^i$.  The goal of this paper is to measure the mean normalized flux as a function of observed wavelength (i.e., redshift) across our entire sample of objects, $F(z) = \langle F^i(z) \rangle$, where $1 + z = \lambda_{\rm obs}/\lambda_{\rm Ly\alpha}$, and $\lambda_{\rm Ly\alpha} = 1215.67$~\AA.  Rather than fit continua to individual spectra, however, we use the fact that the mean continuum for a large group of quasars will be highly similar regardless of redshift.  The difference between composites at different redshifts should therefore primarily reflect the evolution in the mean transmitted flux in the \lya\ forest.  In this section we describe how our composites are constructed, and how these are used to calculate \meanfluxz.

\subsection{Constructing composite spectra}\label{sec:composites}

\begin{table}
   \caption{Composite spectra used in this work.  Columns give the minimum and maximum quasar redshifts, the mean composite redshift, and the number of quasars included in each composite.\vspace{-0.1in}}
   \label{tab:composites}
   \begin{center}
   \begin{tabular*}{8.4cm} {@{\extracolsep{\fill}}cccr}
   \hline
    $z_{\rm min}$  &  $z_{\rm max}$  &  $z_{\rm comp}$  &  $\mathcal{N}$  \\
   \hline
     2.2  &  2.3  &  2.250  &  530  \\
     2.3  &  2.4  &  2.346  &  442  \\
     2.4  &  2.5  &  2.450  &  439  \\
     2.5  &  2.6  &  2.545  &  379  \\
     2.6  &  2.7  &  2.648  &  277  \\
     2.7  &  2.8  &  2.750  &  191  \\
     2.8  &  2.9  &  2.852  &  233  \\
     2.9  &  3.0  &  2.951  &  334  \\
     3.0  &  3.1  &  3.049  &  346  \\
     3.1  &  3.2  &  3.151  &  374  \\
     3.2  &  3.3  &  3.249  &  330  \\
     3.3  &  3.4  &  3.347  &  280  \\
     3.4  &  3.5  &  3.449  &  195  \\
     3.5  &  3.6  &  3.554  &  185  \\
     3.6  &  3.7  &  3.650  &  245  \\
     3.7  &  3.8  &  3.749  &  238  \\
     3.8  &  3.9  &  3.847  &  217  \\
     3.9  &  4.0  &  3.955  &  144  \\
     4.0  &  4.1  &  4.047  &  150  \\
     4.1  &  4.2  &  4.145  &  115  \\
     4.2  &  4.3  &  4.248  &   97  \\
     4.3  &  4.4  &  4.346  &   70  \\
     4.4  &  4.6  &  4.495  &  101  \\
     4.6  &  4.8  &  4.694  &   63  \\
     4.8  &  5.0  &  4.896  &   54  \\
     5.0  &  5.2  &  5.083  &   36   \\
   \hline
   \end{tabular*}
   \end{center}
\end{table}

We constructed 26 composite quasar spectra using a total of 6065 individual spectra from the SDSS DR7 quasar catalogue \citep{schneider2010}.  The redshift range, mean redshift, and the number of quasars contributing to each composite are listed in Table~\ref{tab:composites}.  Redshift bins of $\Delta z = 0.1$ were used between $z = 2.2$ and 4.4.  Above $z = 4.4$ we used a bin size of $\Delta z = 0.2$ in order to increase the number of objects in each composite.  The quasars were selected to have an absolute magnitude $M_{i} > -29.0$, while lacking broad absorption line (BAL) features.  For quasars present in the SDSS DR6 release, BAL quasars were defined using the classifications of \citet{allen2011}, while the small fraction of quasars present only in DR7 were classified via visual inspection.  None of the results presented here, however, are sensitive to the exact definition of the BAL quasar sample.  We also excluded objects where the peak of the \lya\ emission line was greater than ten times the continuum level near a rest wavelength of 1280~\AA.  This was done primarily to avoid very high-redshift objects with extremely low signal-to-noise ratios in the continuum ($S/N \lesssim 2$).  Improved quasar redshifts\footnote{http://www.sdss.org/dr7/products/value\_added/\#quasars} were determined using the \citet{hewett2010} scheme for generating self-consistent redshifts for quasars with SDSS spectra.  We used spectra that had been corrected for telluric emission line residuals in the red \citep{wild2005,wild2010}.  In addition, for roughly 6 per cent of our objects, spectra from multiple epochs were combined to increase the signal-to-noise ratio.

The SDSS spectra are supplied on a logarithmic wavelength scale, with pixels of $69$~\kms\ width. When generating composite spectra in the quasar rest-frame, nearest-pixel values were employed without interpolation or rebinning.  The arithmetic mean over all contributing objects was then computed at each pixel.  All spectra were weighted equally in order to minimize the effects of cosmic variance on the flux in the \lya\ forest.  Error arrays were also computed by directly computing the intra-sample standard deviation.  For the initial composites (described below), pixels falling within 16~\AA \ intervals centered on the wavelengths of strong [\oi] sky emission lines at 5578.5 and 6301.7~\AA\ were masked out.  For the final composites, we masked all regions contaminated by strong telluric lines that are flagged in the bitmask accompanying each SDSS spectrum.  We note that the flagged regions typically do not included all of the pixels in a spectrum affected by skylines, and often cover only the core region contaminated by a skyline and not the line wings, which can still be significantly affected by residuals.  Using spectra that have been corrected for skyline residuals therefore still proved beneficial.

\begin{figure}
   \begin{center}
   \includegraphics[width=0.45\textwidth]{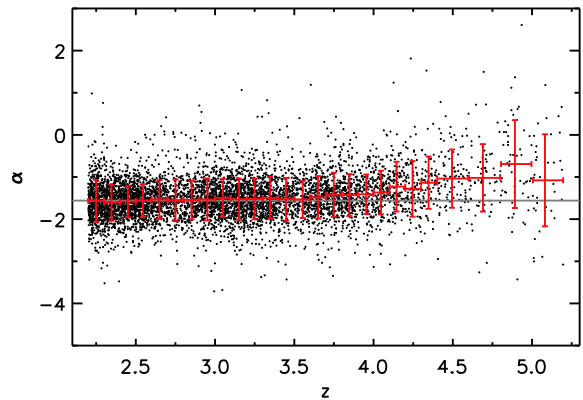}
   \vspace{-0.05in}
   \caption{Continuum power law slopes for the quasars used to construct our composites, where $f_{\lambda} \propto \lambda^{\alpha}$, and $\alpha$ is measured between 1270~\AA\  and 1510~\AA\ (rest wavelengths).  Black points give the values for individual objects.  Red points with error bars show the mean value and standard deviation in each redshift bin.  The horizontal line at $\alpha = -1.56$ shows the composite value from \citet{vandenberk2001}.}
   \label{fig:slopes}
   \end{center}
\end{figure}

\begin{figure*}
   \centering
   \begin{minipage}{\textwidth}
   \begin{center}
   \includegraphics[width=0.85\textwidth]{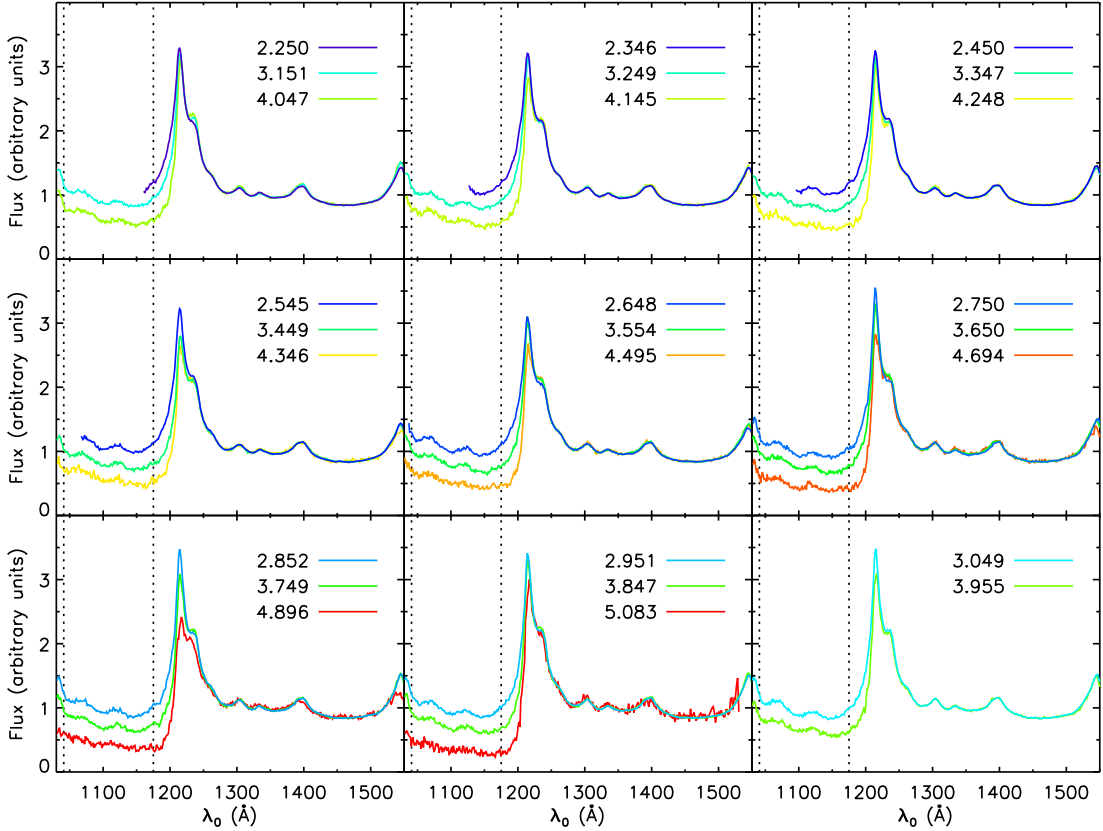}
   \vspace{-0.05in}
   \caption{The twenty-six composite quasar spectra used in this work, plotted versus rest-frame wavelengths.  The spectra are binned using 345~\kms\ pixels, as used in the analysis, and separated into multiple panels for clarity.  The mean quasar redshift is also displayed for each composite.  The vertical dotted lines indicate the region of the \lya\ forest used to measure $F(z)$.  Note that the prominence of the emission lines in the forest naturally decreases with increasing composite redshift as the overall transmitted flux decreases.}
   \label{fig:composites}
   \end{center}
   \end{minipage}
\end{figure*}

Initial composites were generated using the original quasar spectra without modification, except to divide each object by its mean flux near 1285~\AA, which is the region of the continuum closest to the \lya\ forest that is relatively free of emission lines.  These composites exhibited an encouraging similarity in both continuum and emission-line properties longward of 1250~\AA\ in the quasar rest-frame.  A weak but systematic trend, however, appeared in the overall spectral shape as a function of redshift.  To correct for this, we re-calculated the composites after adjusting the continuum slope of each individual object using the following procedure.  The high signal-to-noise ratio initial composite at $z \simeq 2.65$ was chosen as a reference.  For each individual quasar, a power law was then fit to the ratio of the flux in its spectrum to that in the reference composites.  The fit was iterated to reject strongly outlying pixels (e.g., strong metal lines), and the spectrum was divided by the final fit at all wavelengths.  Fits were made between rest-frame wavelengths 1270~\AA\ and 1510~\AA, which is the reddest rest-frame wavelength region covered in the SDSS spectra for objects at $z \sim 5$.  Note that we avoid the high-ionization lines \lya~$\lambda$1216, \nv~$\lambda$1239,1243, and \civ~$\lambda$1548,1551, which tend to vary strongly between objects.  We also exclude the region between 1380~\AA\ and 1410~\AA, which contains the \siiv~$\lambda$1394,1403 emission line doublet.   

The best-fitting spectral indices, calculated by taking the slope of the initial  $z \simeq 2.65$ composite as a reference point, are plotted in Figure~\ref{fig:slopes}.  We note that higher-redshift objects tend to be systematically redder, albeit slightly.   This trend may reflect a combination of effects related both to the selection of the quasars in the SDSS and intrinsic changes in the quasar spectra, or even aspects of the data reduction.  It is clearly essential, however, to remove any systematic trend that could impact the determination of the measurement of the Ly$\alpha$-forest depression with redshift.  The only assumption underlying the nature of the correction procedure adopted is that the essentially featureless changes to the continuum slopes, which were measured over large wavelength intervals redward of the Ly$\alpha$ emission, extend into the spectral region used to study the Ly$\alpha$ forest. We test this assumption in Section~\ref{sec:tests}.

The final composites generated from the continuum-corrected SDSS spectra are shown in Figure~\ref{fig:composites}.  The composite spectra are strikingly similar redward of the \lya+\nv\ emission lines.  The overall amplitudes and slopes are expected to be the same based on the continuum correction procedure described above.  We note, however, that the amplitudes of the emission lines redward of \lya+\nv\ are also very similar.  This is particularly true of the low ionization lines, \oi~$\lambda$1302 + \siii~$\lambda1304$ and \cii~$\lambda$1334, which is encouraging for our purposes since the lines in the forest are also from low ionization species \citep[e.g.][]{vandenberk2001}.

\begin{figure}
   \begin{center}
   \includegraphics[width=0.45\textwidth]{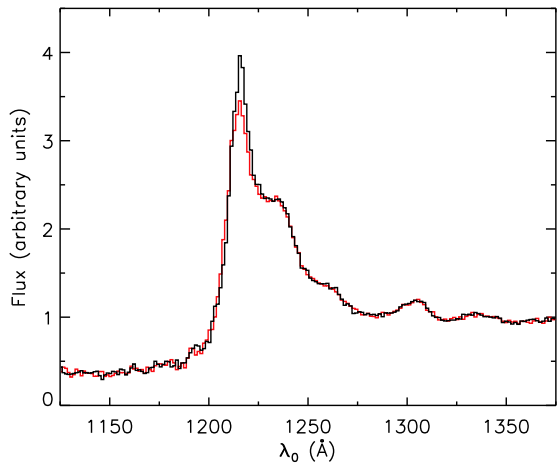}
   \vspace{-0.05in}
   \caption{A demonstration of the effects of redshift errors on composite spectra.  The same set of 56 $4.4 < z < 5.2$ quasar spectra were used to create both versions of the composite plotted here.  The red line shows the composite as generated using individual redshifts measured from the SDSS spectra themselves, while for the black line the redshifts were measured from higher-quality GMOS spectra.  The improved redshifts from the GMOS data produce a more sharply peaked \lya\ emission line but have little effect on the rest of the spectrum.  Note that, since we selected only objects where the GMOS redshifts could be measured from \oi~$\lambda$1302 + \siii~$\lambda1304$ and/or \cii~$\lambda$1334, the emission lines are somewhat stronger than the composites shown in Figure~\ref{fig:composites}.}
   \label{fig:zerr}
   \end{center}
\end{figure}

Significant variations do appear in the strength of the \lya\ emission line, which, taken at face value, might be of concern. The effect, however, is due to errors in the redshifts of the individual objects, which, because of \lya\ absorption on the blue side of the \lya\ emission line, tend to decrease the height of the emission in the composite spectrum. The effect increases towards higher redshift as the absorption in the \lya\ forest increases.  The redshift uncertainties for the SDSS quasar spectra also increase with redshift \citep[Table 3 of][]{hewett2010} as the \civ\ emission line has to be used in the redshift determination. At the very highest redshift, $z_{\rm comp} \simeq 5.08$, the effect apparently diminishes (see Figure~\ref{fig:composites}) because the Ly$\alpha$ emission line itself dominates in the redshift determination, causing the peak of the composite Ly$\alpha$ line to sharpen by selection.  By comparison, the flux over the region spanning $1250\,{\mbox \AA} < \lambda < 1500\,{\mbox \AA}$, which contains relatively weak emission lines, shows very little variation with redshift.  Since the emission lines in the forest are also weak, redshift errors should have a minimal effect there, as well.  

We directly verified that redshift uncertainties affect the strength of \lya\ by comparing composite spectra of 56 quasars at $4.4 < z < 5.2$ with improved redshifts from recently-obtained Gemini/GMOS spectra (Worseck et al, in prep).  Two composites were generated using the same set of SDSS spectra.  In one case we use the redshifts measured directly from the SDSS spectra, while in the other case we use redshifts from \oi~$\lambda$1302 + \siii~$\lambda1304$ and/or \cii~$\lambda$1334 measured from the GMOS data.  The results are shown in Figure~\ref{fig:zerr}.  The improved GMOS redshifts produce a stronger and more sharply peaked \lya\ line, as expected, while the remainder of the spectrum is nearly unaffected.  This supports our assumption that, for all of our composites, the  differences in the fluxes blueward of the \lya\ emission lines are due to the evolution of the mean opacity of the \lya\ forest.   It is these differences that we will use to constrain \meanfluxz.

\subsection{Calculating the mean transmitted flux}\label{sec:method}

At a given rest wavelength $\lambda_0$, the flux of a quasar composite with mean emission redshift $z_{\rm comp}$ will be attenuated by a factor 
\begin{equation}\label{eq:flux}
F(z) = \frac{f_{\rm obs}(\lambda_0,z_{\rm comp})}{f_{\rm cont}(\lambda_0)} \, , 
\end{equation}
where $1+z = \lambda_0 (1+z_{\rm comp})/\lambda_{\rm Ly\alpha}$.  We unfortunately do not have a completely unabsorbed composite at $z_{\rm comp} = 0$ that would allow us to measure $f_{\rm cont}(\lambda_0)$ absolutely, and hence calculate $F(z)$ directly.  However, we can measure the ratio of $F(z)$ at two redshifts by evaluating the ratio of fluxes between two composites at the same rest-frame wavelengths,
\begin{equation}
\frac{F(z_1)}{F(z_2)} = \frac{f_{\rm obs}(\lambda_0,z_{\rm comp,1})}{f_{\rm obs}(\lambda_0,z_{\rm comp,2})}  \, .
\end{equation}
Here, the dependence on the continuum has been eliminated because we are using composites rather than individual spectra.  This relation will hold to the degree that the mean continuum in the composites either does not evolve or has been adequately corrected with redshift, an assumption we test in Section~\ref{sec:results}.  Nominally, however, this dataset allows us to compute $F(z)$ up to a normalization factor by finding a function that satisfies the set of observed flux ratios.

We parametrized $F(z)$ as a discrete function in bins of $\Delta z = 0.1$, with the first bin centered on $z = 2.15$.  We then computed $F(z)$ as a fraction of the transmitted flux in that bin, i.e., $F(z)/F(z=2.15)$.  Fits to $F(z)$ were computed using an IDL implementation of the Levenberg-Marquardt least-squares algorithm \citep{markwardt2009}.  In order to avoid the quasar proximity region at the red end of the forest and contamination from \ovi~$\lambda$1032,1038 absorption at the blue end, we restrict the fits to rest-frame wavelengths $1041\,{\mbox \AA} < \lambda_0 < 1175\,{\mbox \AA}$.   When analyzing the data, we bin the composite fluxes into 345~\kms\ pixels.  This is to avoid large non-symmetric errors in $F(z_1)/F(z_2)$ when $F(z_2)$ becomes small, although nearly identical results were obtained without binning.  We also made secondary adjustments to the composite continua by dividing each composite by a power-law fit over $1270\,{\mbox \AA} < \lambda_0 < 1510\,{\mbox \AA}$ to the ratio of that composite to the average of  all composites with $2.25 < z_{\rm comp} <  4.25$.  This was a minor correction to help remove any residual differences in the composite shapes, and changed the final $F(z)$ values by less than the 1$\sigma$ errors.

The primary goal of this paper is to measure the mean opacity of intergalactic hydrogen to \lya.  Here we loosely define ``intergalactic'' to include all diffuse, highly-ionized absorbers with \hi\ column densities $N_{\rm H\,I} < 10^{19}~{\rm cm^{-2}}$.  As defined in equation~\ref{eq:flux}, however, $F(z)$ in the \lya\ forest will also include absorption from metal lines, as well as from higher-column density \lya\ absorbers.  We now describe how we account for these additional contributions to arrive at the desired $F(z)$ values.

\subsubsection{Metal lines}\label{sec:metals}

Absorption in the \lya\ forest due to intervening metals, while small compared to the \lya\ absorption, must be accounted for in order to measure $F(z)$ precisely.  Conventional studies using continuum-normalized spectra have relied on masking \citep[e.g.,][]{schaye2003} or fitting out \citep{kim2007} metal lines individually, or else applying global statistical corrections \citep[e.g.,][]{kirkman2005}.   Measuring flux ratios in composite spectra, however, offers an alternative approach to mitigating the impact of metals, as discussed below.

The mean level of metal-line absorption may vary, in principle, with both quasar redshift and rest-frame wavelength.  A large fraction of absorption in the forest will come from the \civ~$\lambda1548,1551$, \siiv~$\lambda1394,1403$,  and \mgii~$\lambda2796,2804$ doublets, which, because they have rest wavelengths longward of \lya, will produce, on average, a continuum of absorption extending over the region redward of the \lya\ emission line we use to scale our spectra and into the forest.  An additional contribution will come from the various metal lines associated with high \hi\ column density systems such as DLAs.  These lines have a variety of rest wavelengths, including some that fall between \lyb\ and \lya.

In practice, the flux decrement from metals is relatively minor, and appears to depend only mildly on either quasar redshift or rest-frame wavelength.  \citet{becker2011a} found a nearly constant decrement of $\sim$3 per cent in the region between the \nv~$\lambda1239,1243$ and \siiv~$\lambda1394,1403$ emission lines in the high-resolution spectra of a large sample of quasars with redshifts  $z \sim 2-4$.  \citet{kirkman2005} also found that the absorption due to metals redward of \lya\ evolves slowly with quasar redshift.  In the \lya\ forest, \citet{schaye2003} found a similar ($\sim$3 percent) level of absorption from metals over \lya\ redshifts $z \sim 2-4$, with little evidence for strong evolution.  \citet{kim2007}, in possibly the most careful such work to date,  also found a mean decrement of $\sim$3 per cent due to all metals in the forest over $z \sim 2-3$, albeit with significant scatter between sight lines, and no evidence for evolution.

In a composite spectrum, the absorption due to metals will essentially scale the mean flux by a multiplicative factor, and may also introduce a change in slope with rest-frame wavelength.  Given the apparently mild dependence of this absorption on either rest-frame wavelength or quasar redshift, however, we expect these effects to be minor (on the order of a few percent).  More importantly, the process of scaling the composites to have the same amplitude and slope between the \lya\ and \civ\ emission lines will tend to eliminate the relative differences in the metal line absorption as a function of composite redshift.  This is another advantage of our differential approach to measuring $F(z)$.  Systematic errors with redshift may remain if the metal-line absorption (or more precisely, the relative difference in metal-line absorption between composites at different redshifts) is not well represented by a power law in wavelength (see Section~\ref{sec:composites}).  Given the metal-line measurements discussed above, however, we expect such residual systematics to be minor, even compared to relatively small statistical errors on $F(z)$ from such a large data set.  

We note that establishing an absolute scaling for $F(z)$ still requires a correction for metal lines, at least over a narrow range in redshift.  As described below, we use absolute values of $F(z \sim 2)$ from \citet{fg2008a} that have been corrected using measurements from \citet{schaye2003}.  This correction for metal lines is roughly consistent with other works, although it is still a potential source of systematic error in our final $F(z)$ values, albeit at the $\lesssim$1 per cent level.  For example, using the \fg\ values corrected for metals following \citet{kirkman2005} would give absolute values for $F(z)$ that are 0.5 per cent higher.

\subsubsection{Correction for optically thick absorbers}

The total \lya\ opacity will include a contribution from absorbers that are optically thick to ionizing photons (i.e., with $N_{\rm H\, I} \ge 10^{17.2}~{\rm cm^{-2}}$).  As discussed below, most of this absorption will come from super Lyman limit systems (SLLSs; $10^{19.0}\,{\rm cm^{-2}} \le N_{\rm H\, I} < 10^{20.3}\,{\rm cm^{-2}}$) and damped \lya\ absorbers (DLAs; $N_{\rm H\, I} \ge 10^{20.3}\,{\rm cm^{-2}}$).  These are believed to be associated with intervening galaxies, and are difficult to model in numerical simulations.  In order to facilitate a direct comparison with simulations, therefore, we correct for these absorbers in order to arrive at our final $F(z)$ values.   Following \citet{fg2008a}, we use a model to calculate the integrated absorption due to optically thick systems.  For a given frequency distribution in column density, $N_{\rm H\, I}$, Doppler parameter, $b$, and redshift, the flux decrement due to these absorbers will be
\begin{equation}
1 - F_{\rm T}(z) = \frac{1+z}{\lambda_{\rm Ly\alpha}} \int_{N_{\rm min}}^{N_{\rm max}} dN_{\rm H\,I} \int db \, f(N_{\rm H I}, b, z) \, W_0(N_{\rm H I}, b) \, ,
\end{equation}
where $W_0$ is the rest-frame equivalent width.  At high column densities, $W_0$ will depend mainly on the column density, and so we set $b = 20$~\kms, as in \citet{fg2008a}.  We assume that the shape of the column density distribution does not evolve with redshift, such that $f(N_{\rm H\,I}, z)  = f(N_{\rm H\,I}) \, dn/dz$.  We use the broken power-law column density distribution at $z \sim 3.7$ compiled by \citet{prochaska2010}, and adopt $10^{19}\,{\rm cm^{-2}}$ and $10^{22}\,{\rm cm^{-2}}$ as our limits of integration.  Lyman limit systems with $10^{17.2}\,{\rm cm^{-2}} \le N_{\rm H\, I} < 10^{19.0}\,{\rm cm^{-2}}$ were found to contribute a negligible amount of \lya\ absorption at our level of precision, while DLAs with $N_{\rm H\,I} > 10^{22}\,{\rm cm^{-2}}$ are vanishingly rare.  The redshift evolution of the number density of SLLSs and DLAs is not known precisely over the entire redshift range; however, $dn/dz \propto (1+z)^2$ gives a reasonable fit to the evolution of  DLAs over $2.3 < z < 4.4$ measured by \citet{prochaska2009a}.  Adopting this evolution gives an integrated flux decrement due to SLLSs and DLAs of $1 - F_{\rm T}(z) = 0.0045 \, [(1+z)/3]^{3.0}$.  We note that this is somewhat more absorption than estimated by \fg, who adopted a model with $f(N_{\rm H\,I}) \propto N_{\rm H\,I}^{-1.5}$ over all column densities, and $dn/dz \propto (1+z)^{1.5}$.

When calculating $F(z)$, we first fit raw values relative to $z \sim 2$ that include absorption from both the optically thin \lya\ forest and from optically thick absorbers.  We then corrected these relative values using the above scaling for $F_{\rm T}$ to give the relative mean flux in the forest only.  These corrected values of $F(z)/F(z=2.15)$ were then scaled to measurements at $z \le 2.5$ from the literature to derive absolute values for $F(z)$, as described below.

\subsubsection{Absolute scaling}

In order to compute absolute values for $F(z)$, we scaled our results to $F(z)$ measurements made from high-resolution spectra at redshifts where continuum uncertainties are minimal.  We used the results of \citet{fg2008a}, who used artificial spectra drawn from cosmological simulations to correct for continuum-fitting biases.  These authors found the corrections to be small ($\lesssim 2$~per cent) at $z < 2.5$.  Their results were also confirmed by \citet{becker2011a}, who measured $F(z)$ from high-resolution spectra using a different approach to correct for continuum errors.  Becker et al. re-normalized both the real data and the simulated spectra by deliberately placing the continuum along the tops of the highest transmission peaks.  The true $F(z)$ was then measured from the simulations once the simulated optical depths had been adjusted such that the re-normalized data and simulations had the same mean flux.  The resulting $F(z)$ agreed with the \fg\ measurements to within one per cent over $2 < z < 2.5$.  We note that \fg\ and Becker et al. also used different methods to correct for metal-line contamination; the \fg\ values adopted here include a correction based on metal-line measurements in the forest from \citet{schaye2003}, whereas Becker et al. applied a similar correction based on absorption redward of \lya.

To scale our measurements, we interpolated our $F(z)/F(z=2.15)$ results onto the centers of the four $\Delta z = 0.1$ redshift bins between $z = 2.2$ and 2.5 used by \citet{fg2008a}.  The mean flux ratio between our raw results and the \fg\ values, weighted by the errors, was then used to compute our final $F(z)$.  We note that when computing the scaling we first adjusted the \fg\ data to account for the difference in the amount of absorption we assume for optically thick absorbers.  This is a minor ($\sim$0.1 per cent) correction at $z \le 2.5$, however.

\begin{figure*}
   \centering
   \begin{minipage}{\textwidth}
   \begin{center}
   \includegraphics[width=1.00\textwidth]{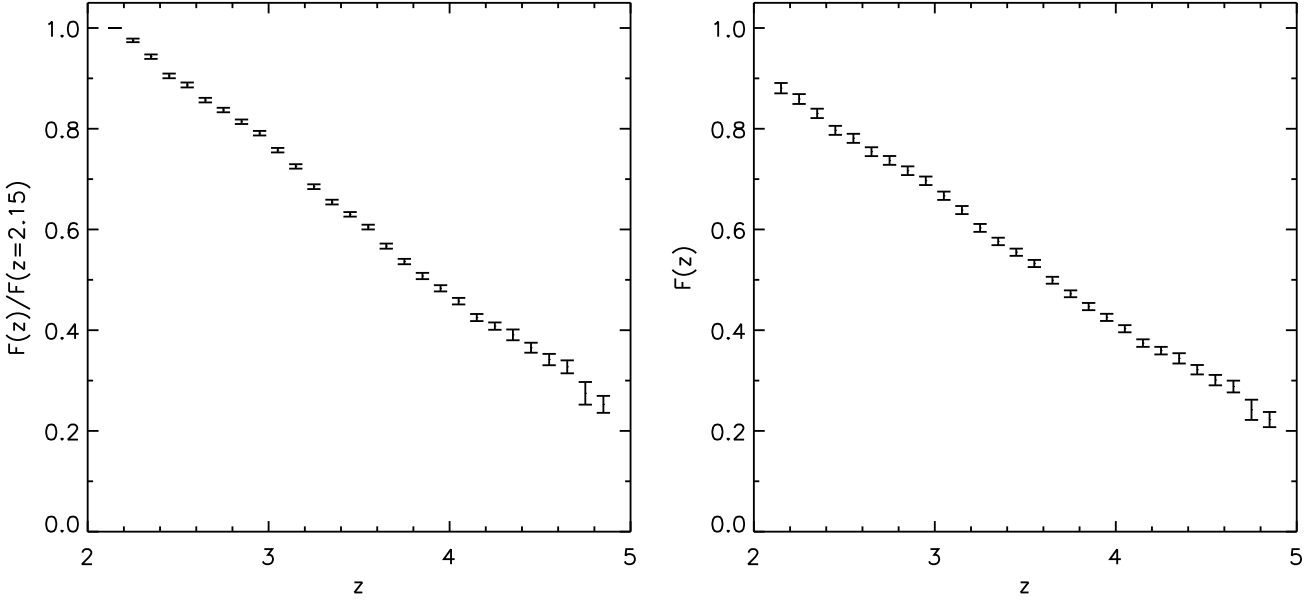}
   \vspace{-0.05in}
   \caption{The mean transmitted flux in the \lya\ forest.  Left panel: $F(z)$ as a fraction of the mean transmitted flux at $z = 2.15$.  These values are computed directly from our composite spectra, and include a correction for the absorption due to optically thick systems.  Right panel: $F(z)$ values after scaling our results to the measurements of \citet{fg2008a} over $2.2 \le z \le 2.5$.  Error bars are 1$\sigma$, and include uncertainties in the scaling factor for $F(z)$ points in the right-hand panel.  Tabulated values are given in Table~\ref{tab:meanflux}.  The full covariance matrices can be found in the accompanying supplemental material.}
   \label{fig:meanflux}
   \end{center}
   \end{minipage}
\end{figure*}

\begin{figure*}
   \centering
   \begin{minipage}{\textwidth}
   \begin{center}
   \includegraphics[width=0.70\textwidth]{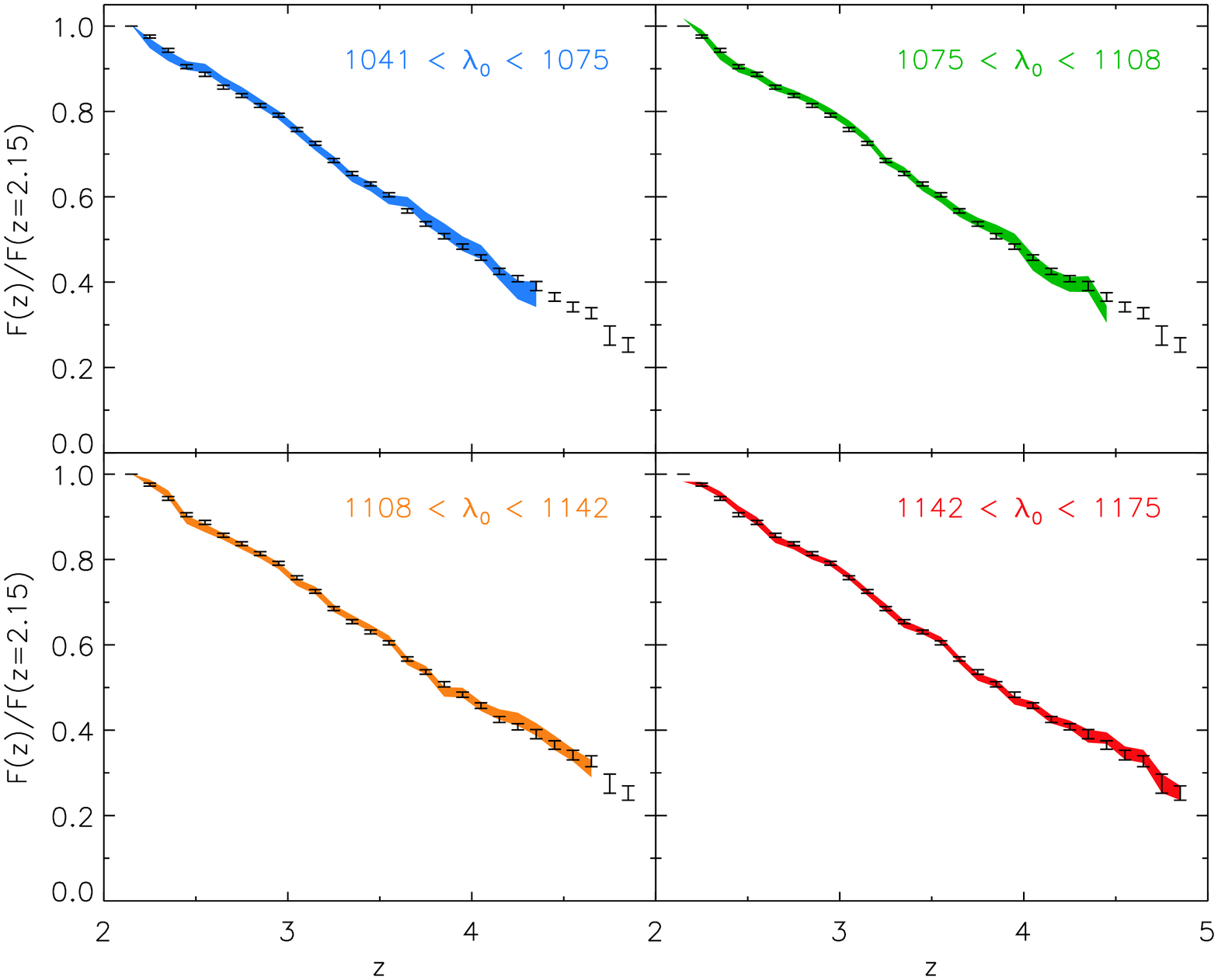}
   \vspace{-0.05in}
   \caption{The dependence of the recovered mean transmitted flux on the rest-frame wavelength of the \lya\ forest used to measure $F(z)$.  In each panel, discreet points with error bars show the reference values of $F(z)/F(z=2.15)$ calculated using the entire forest (left-hand panel of Figure~\ref{fig:meanflux}).  Shaded regions show the results using only the indicated rest-frame wavelength interval.  The values for each interval have been slightly scaled to match those from the entire forest at redshifts where they overlap.  This scaling reflects small variations in the relative value of $F(z=2.15)$, and does not affect the overall shape of the mean flux evolution with redshift.  Both the discreet error bars and the shaded error regions are 1$\sigma$.  Errors in adjacent redshift bins are again correlated.  Systematic differences in the composite continua would be expected to give rise to different results for $F(z)$ when measured from different regions of the forest.  The fact that the shape of $F(z) $ does not change significantly with rest-frame wavelength suggests that such systematic errors are within the bootstrap uncertainty estimates.}
   \label{fig:sections}
   \end{center}
   \end{minipage}
\end{figure*}

\subsubsection{Error estimates}

Uncertainties in $F(z)$ were estimated using a bootstrap resampling approach.  In each iteration, a new composite was generated for each redshift bin from a set of $\mathcal{N}$ quasars drawn randomly, with replacement, from the set of $\mathcal{N}$ objects in that bin.  $F(z)$ was then re-computed from the new composites following the procedure described in Section~\ref{sec:composites}.  This process was repeated 1000 times, and the full covariance arrays for both the unscaled $F(z)/F(z=2.15)$ and scaled $F(z)$ were computed from the results.  We note that the bootstrap errors are an order of magnitude larger than the formal errors in $F(z)$ computed (but not used further) from the least-squares fit.  This suggests that statistical uncertainty in the composite SEDs dominates over the intrinsic sightline-to-sightline variance in $F(z)$, which is incorporated into the formal variance array when constructing each composite.  We also stress that, because we are fundamentally fitting ratios of fluxes at different redshifts, the errors in $F(z)$ will be correlated.  Any analysis that uses the $F(z)$ results should therefore consider the entire covariance array.

\subsubsection{Impact of sky-subtraction errors}

Measuring $F(z)$ at $z < 2.3$ requires analyzing fluxes at very blue wavelengths ($\lambda_{\rm obs} < 4000$~\AA).  Recent papers using SDSS data \citep{prochaska2010,paris2011}, however, have noted potential problems with the sky-subtraction in SDSS spectra at these wavelengths.  \paris, for example, found a net positive residual below 4000~\AA\ in a sample of quasar spectra that should show no flux below 4280~\AA\ due to the presence of a DLA at $z > 3.7$.  In principle, a positive residual would cause us to overestimate $F(z)$ at $z < 2.3$, and would also impact our higher-redshift measurements, since the errors in our $F(z)$ points will be correlated.  We note that, for our measurements, the impact of sky-subtraction residuals in the far blue is likely to be small; we only use data at $\lambda_{\rm obs} < 4000$~\AA\ in the lowest-redshift composites ($z_{\rm comp} \le 2.75$), and these were constructed using relatively bright quasars.  Nevertheless, we performed multiple tests to determine whether sky-subtraction errors could significantly affect our $F(z)$ results.

We first directly estimated the mean sky residual (and any contamination from intervening objects within the SDSS fibers) by measuring the flux over $\lambda_{\rm obs} = 3900-4000$~\AA\ in the spectra of each of our  quasars at $z > 4.5$.  These objects should generally not have any flux below 4000~\AA\ due to the integrated Lyman-series opacity of the IGM and any optically thick absorbers.  A few exceptionally clear lines of sight may show some transmission, in which case the flux will be an upper limit on the sky-subtraction residuals.  We measured a mean flux of $f_{\rm resid}^{4000} \simeq (8 \pm 1) \times 10^{-19}~{\rm erg\,s^{-1}\,cm^{-2}\,\AA^{-1}}$.  For our quasars at $z < 2.8$, this is, on average, roughly 0.5 per cent of the observed flux over the same wavelength range, i.e., $\langle f_{\rm resid}^{4000} / f_{\rm obs}^{4000} \rangle \simeq 0.005$.  Since we are making a differential measurement, subtracting this value of $f_{\rm resid}^{4000}$ from all spectra at $\lambda < 4000$~\AA\ prior to constructing the composites would therefore affect the shape of our $F(z)$ result by increasing the values of $F(z)/F(z=2.15)$ at $z > 2.3$ by 0.5 per cent.  This is comparable to or less than the 1$\sigma$ errors at $z > 2.3$, as described below.  Since the absolute scaling is performed over redshifts that include observed wavelengths greater than 4000~\AA, the final scaled $F(z)$ values would be less affected (0.03 percent lower at $z < 2.3$ and 0.02 percent higher at $z > 2.3$ in this simplified model for the sky residuals).  This further supports the conclusion that, since the quasars we use to measure the flux below 4000~\AA\ are bright, sky-subtraction errors at these wavelengths will have a relatively minor impact.

Second, we tried adjusting the redshift interval over which we scaled our absolute $F(z)$ values.  If sky-subtraction errors are artificially increasing our $F(z)/F(z=2.15)$ values at $z < 2.3$, the matching our results to literature values at these redshifts should yield a normalization factor that is too low.  To test this, we tried scaling our results to the \citet{fg2008a} measurements over $2.6 \le z \le 3.0$, rather than over $2.2 \le z \le 2.5$.  The resulting scale factor was higher by 0.5 per cent, which is well within the 1$\sigma$ uncertainty of $\sim$1 per cent (see measurements below).

Finally, we tried fitting $F(z)$ using only data at absorption redshifts $z > 2.5$, or $\lambda_{\rm obs} > 4250$~\AA.  This had a negligible impact on the shape of $F(z)$ at $z > 2.5$.  In summary, therefore, sky-subtraction residuals in the far blue do not appear to have a significant impact on our results.  We emphasize, however, that systematics such as these may need to be more carefully addressed in order to achieve substantially higher-precision mean flux measurements than the ones presented here.

\section{Results}\label{sec:results}

Our results for $F(z)/F(z=2.15)$, along with our $F(z)$ values scaled to the \citet{fg2008a} measurements at $z \le 2.5$, are shown in Figure~\ref{fig:meanflux} and listed in Table~\ref{tab:meanflux}.    The errors are 1$\sigma$ and reflect only the diagonal elements of the covariance matrix.  For $F(z)$ the errors include the uncertainty in the overall absolute scaling, which has a fractional uncertainty of $\sim$1 per cent (for the adopted set of corrections for continuum bias, metal-line contamination, and other systematics made by \fg)     and dominates the error budget at $z \le 3.75$.  

The mean flux exhibits a smooth decline with redshift over $2 < z < 5$.  Notably, we see no indication of the feature at $z \sim 3.2$ reported by \citet{bernardi2003}, \citet{fg2008a}, and \citet{dallaglio2008}.  Comparisons with literature values are discussed in Section~\ref{sec:comparison}.

\subsection{Tests for variations in the intrinsic composite quasar continua}\label{sec:tests}

Our primary systematic uncertainty is whether the shape and amplitude of the mean unabsorbed continuum in the \lya\ forest is the same for all composites, regardless of redshift.  We naturally expect some variation in the mean quasar continuum with redshift due to sample variance, redshift-dependent color-selection techniques, and/or genuine evolution in the mean continuum shape over $1040\,{\mbox \AA} < \lambda_0 < 1500\,{\mbox \AA}$.  We emphasize, however, that our technique does not require the mean uncorrected quasar SED to be the same in each emission redshift bin.  Instead, the composites must only have the same unabsorbed SED once the individual quasar continua are corrected following the procedure described in Section~\ref{sec:composites}.  In the following two tests we address whether variations in the quasar SEDs have been adequately accounted for, and hence whether residual differences in the composite continua may be significantly affecting our $F(z)$ results.

We first computed the unscaled $F(z)$ using different regions of the \lya\ forest.  As described in Section~\ref{sec:composites}, the continuum slopes for individual quasars are adjusted using power-law fits made redward of the \lya\ emission line that are then extrapolated over the forest.  We naturally expect the accuracy of these fits to diverge with increasing wavelength offset from the fitted region.  If systematic differences remain in the mean composite continua over the forest as a result of errors in the fits, therefore, they should be more pronounced at bluer wavelengths.  In addition, although power-law corrections produce highly similar composites redward of \lya, there may be additional variations with redshift in the mean quasar continuum over the forest that are not adequately captured by a power law.  In either case, if significant variations exist in the composite continua, then computing $F(z)$ from different regions of the forest should produce systematically different results.  To test this, we divided the \lya\ forest from all composites into four quartiles in rest-frame wavelength.  Figure~\ref{fig:sections} shows $F(z)$ computed from each quartile after scaling the results to match the mean amplitude of $F(z)/F(z=2.15)$ computed from the entire forest over redshift bins where the results overlap.  The scaling reflects small differences in the relative value of $F(z=2.15)$ for each quartile, and does not affect the overall shape of $F(z)$.  For all quartiles, the derived mean flux is consistent with that measured from the entire forest to within the (correlated) errors.  The fact that significant differences in the shape of $F(z)$ do not appear when using different rest-frame wavelengths indicates that systematic errors in $F(z)$ related either to errors in the power-law extrapolations or to additional intrinsic variations in the composite continua with redshift are minimal and within the error bars estimated from bootstrap resampling.

We next checked for systematic residuals in the composite spectra after subtracting observed flux models computed from our measured $F(z)$ and the inferred mean continuum over the \lya\ forest.  Part of the robustness of this approach to determining $F(z)$ is that, for all redshifts $z \le 4.7$, a given absorption redshift is covered by multiple composites, each at a different rest-frame wavelength.  If the assumption that the intrinsic continuum is the same for all composites is incorrect, therefore, the residuals from the best-fit models should show systematic deviations where the composites overlap in absorption redshift (observed wavelength).  For example, if the intrinsic continua over the forest get progressively bluer (redder) with increasing composite redshift, the residuals would show a systematically negative (positive) slope.  Smaller-scale differences in the continua would similarly produce opposite-signed residuals in overlapping composites.

\begin{figure}
   \begin{center}
   \includegraphics[width=0.45\textwidth]{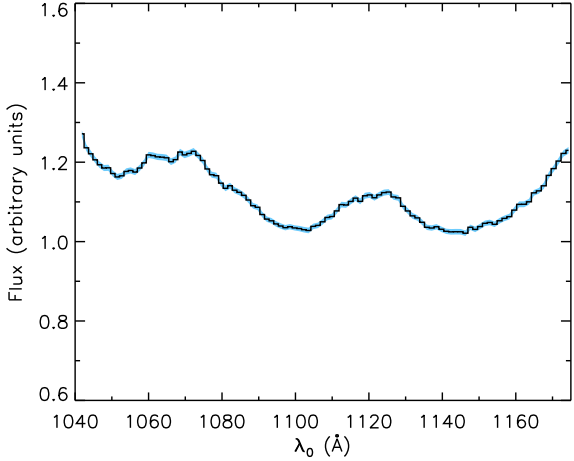}
   \vspace{-0.05in}
   \caption{The mean quasar continuum in the \lya\ forest.  The continuum was calculated by averaging the composites with $z_{\rm comp} < 4.05$ after correcting each one for the measured relative mean transmitted flux, $F(z)/F(z=2.15)$, calculated separately at each observed wavelength.  The overall amplitude of the continuum has therefore implicitly been multiplied by $F(z=2.15)$.  The solid line gives the results from the full sample.  The shaded region shows the 1$\sigma$ uncertainties calculated from bootstrap trials.}
   \label{fig:continuum}
   \end{center}
\end{figure}

We reconstructed the continuum in the forest for each composite, scaled by $F(z=2.15)$, by dividing the observed fluxes by our raw (uncorrected for absorption due to optically thick systems) relative transmitted flux values, 
\begin{equation}
f_{\rm cont}(\lambda_0,z_{\rm comp}) F(z=2.15) = \frac{f_{\rm obs}(\lambda_0,z_{\rm comp})}{F(z)/F(z=2.15)} \, .
\end{equation}
Here, $z$ is calculated from the observed wavelength of each pixel, and the transmitted flux is interpolated between the values plotted in Figure~\ref{fig:meanflux}.  Observed wavelengths for a composite at redshift $z_{\rm comp}$ are related to the rest-frame wavelengths plotted in Figure~\ref{fig:composites} simply by $\lambda_{\rm obs} = \lambda_0(1+z_{\rm comp})$.   We then computed the mean scaled continuum across all composites with $z_{\rm comp} < 4.05$, where the redshift limit was chosen to avoid noisier composites at higher redshifts.  The result is shown in Figure~\ref{fig:continuum}.  Finally, a model for the observed flux for each composite was created by multiplying the mean scaled continuum, $f_{\rm cont}(\lambda_0)F(z=2.15)$, by the relative mean transmitted flux, $F(z)/F(z=2.15)$, evaluated separately for each pixel.

\begin{figure*}
   \centering
   \begin{minipage}{\textwidth}
   \begin{center}
   \includegraphics[width=1.00\textwidth]{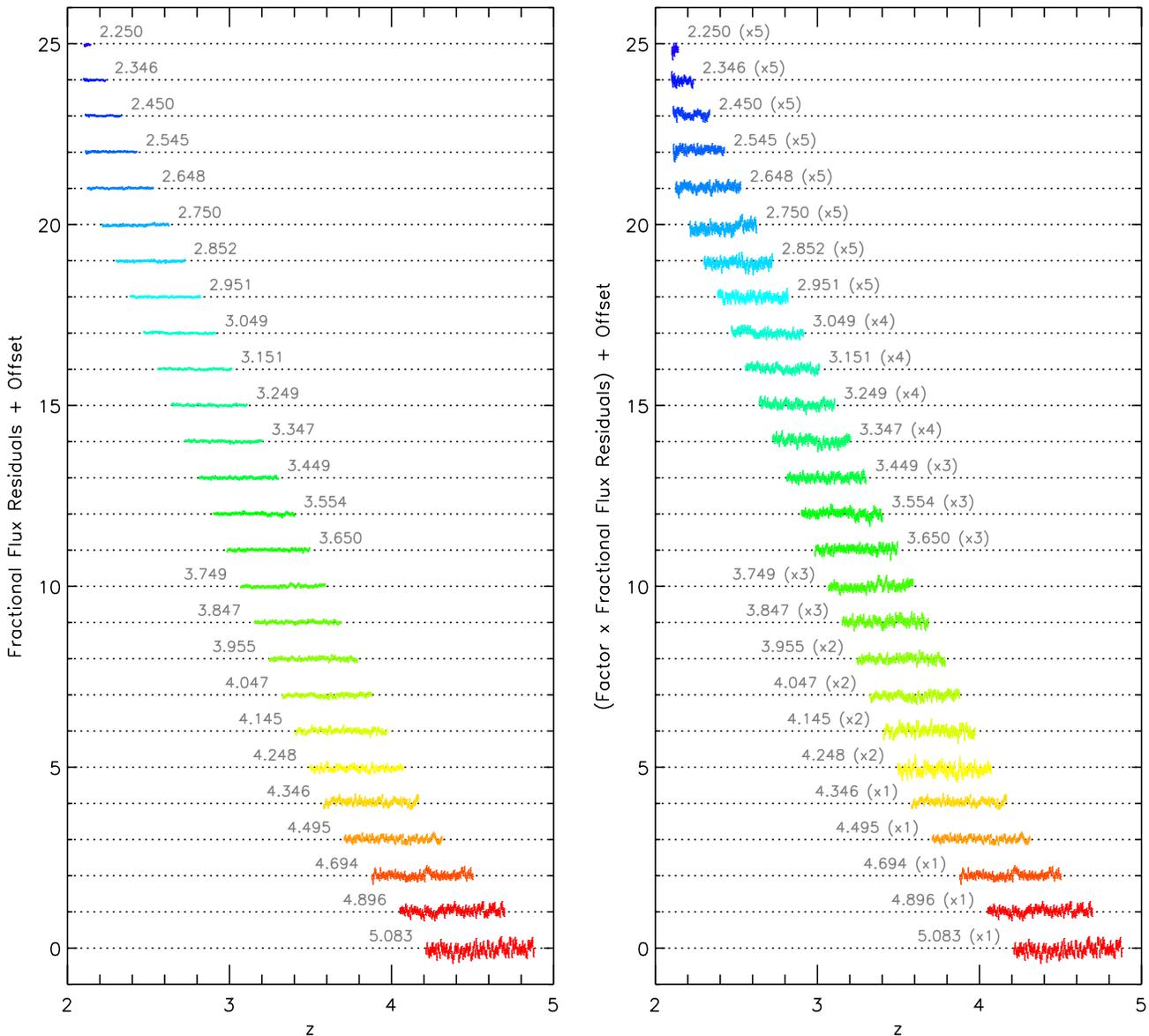}
   \vspace{-0.05in}
   \caption{Fractional flux residuals in the \lya\ forest for each of the twenty-six composites, plotted with respect to the corresponding \lya\ absorption redshifts.  The residuals are calculated as $(f_{\rm obs}-f_{\rm model})/f_{\rm model}$, where $f_{\rm model}$ is the product of the mean continuum shown in Figure~\ref{fig:continuum} and the relative mean transmitted flux, $F(z)/F(z=2.15)$, evaluated at each observed wavelength (i.e., redshift).  1$\sigma$ error bars were computed from bootstrap trials.  A unity offset is added between composites for clarity.  The redshift of the composite is printed next to each line.  In the left-hand panel, the residuals are plotted at their original amplitudes.  In the right-hand panel, the residuals have been multiplied by the factors in parentheses.  The fact that the residuals generally lie within the 1$\sigma$ error ranges indicates that the  continuum over the \lya\ forest (i.e., the slope of the underlying power-law continuum and the relative shape and amplitude of the emission lines) changes little with quasar redshift, and that, to within the present errors, the ratios of observed fluxes between composites can be well modeled by the evolution in the mean \lya\ forest opacity.}
   \label{fig:residuals}
   \end{center}
   \end{minipage}
\end{figure*}

The fractional residuals from these models, $(f_{\rm obs} - f_{\rm model})/f_{\rm model}$, are shown in Figure~\ref{fig:residuals}.  Error bars were computed by repeating the above procedure for each bootstrap trial.  While small offsets do appear for some composites, the residuals are generally within the 1$\sigma$ error ranges.  This suggests that, at least to within the present error estimates, the unabsorbed continua are sufficiently similar between composites that the ratios of observed fluxes at the same rest-frame wavelengths can be adequately modeled by the evolution in $F(z)$ plus a minor contribution from optically thick absorbers.  

We emphasize that these tests address the self-similarity of the continuum only over the \lya\ forest.  Variations with redshift in the observed mean quasar SED at other wavelengths are known to exist.  For example, \citet{worseck2011b} found that SDSS, in order to avoid contamination from the stellar locus, preferentially selects quasars with red SEDs in the $u$ and $g$ bands over $2.7 \lesssim z \lesssim 3.5$.  They found that this primarily effects the observed mean flux for higher-order Lyman-series transitions, and has relatively little impact on the \lya\ forest (see their Figure 17).  Nevertheless, careful modeling of such selection effects may be required in order to achieve significantly more precise mean flux measurements than the ones presented here.

\begin{table}
   \caption{Tabulated values for the mean transmitted flux in the \lya\ forest.  Columns give the redshift of each bin, the mean transmitted flux as a fraction of its value at $z = 2.15$ before and after correcting for absorption due to optically-thick absorbers, and the corrected absolute mean transmitted flux after scaling to the measurements of \citet{fg2008a} over $2.2 \le z \le 2.5$. The full covariance matrices can be found in the accompanying supplemental material.\vspace{-0.1in}}  
   \label{tab:meanflux}
   \begin{center}
   \begin{tabular*}{8.4cm} {@{\extracolsep{\fill}}cccc}
   \hline
    $z$  &  \multicolumn{2}{c}{$F(z)/F(z=2.15)$}            &  $F(z)$ \\
         &  Raw               &  Corrected  		      &         \\
   \hline
     2.15  &  $1.0000$             &  $1.0000$             &  $0.8806 \pm 0.0103$  \\
     2.25  &  $0.9750 \pm 0.0035$  &  $0.9755 \pm 0.0035$  &  $0.8590 \pm 0.0098$  \\
     2.35  &  $0.9421 \pm 0.0044$  &  $0.9431 \pm 0.0044$  &  $0.8304 \pm 0.0093$  \\
     2.45  &  $0.9034 \pm 0.0046$  &  $0.9049 \pm 0.0046$  &  $0.7968 \pm 0.0089$  \\
     2.55  &  $0.8849 \pm 0.0049$  &  $0.8869 \pm 0.0049$  &  $0.7810 \pm 0.0090$  \\
     2.65  &  $0.8543 \pm 0.0045$  &  $0.8568 \pm 0.0045$  &  $0.7545 \pm 0.0088$  \\
     2.75  &  $0.8341 \pm 0.0044$  &  $0.8371 \pm 0.0044$  &  $0.7371 \pm 0.0088$  \\
     2.85  &  $0.8104 \pm 0.0044$  &  $0.8139 \pm 0.0044$  &  $0.7167 \pm 0.0086$  \\
     2.95  &  $0.7871 \pm 0.0045$  &  $0.7911 \pm 0.0045$  &  $0.6966 \pm 0.0084$  \\
     3.05  &  $0.7530 \pm 0.0045$  &  $0.7574 \pm 0.0045$  &  $0.6670 \pm 0.0082$  \\
     3.15  &  $0.7203 \pm 0.0044$  &  $0.7252 \pm 0.0044$  &  $0.6385 \pm 0.0080$  \\
     3.25  &  $0.6797 \pm 0.0047$  &  $0.6849 \pm 0.0047$  &  $0.6031 \pm 0.0079$  \\
     3.35  &  $0.6487 \pm 0.0047$  &  $0.6543 \pm 0.0047$  &  $0.5762 \pm 0.0074$  \\
     3.45  &  $0.6241 \pm 0.0045$  &  $0.6301 \pm 0.0045$  &  $0.5548 \pm 0.0071$  \\
     3.55  &  $0.5983 \pm 0.0046$  &  $0.6047 \pm 0.0047$  &  $0.5325 \pm 0.0071$  \\
     3.65  &  $0.5604 \pm 0.0050$  &  $0.5669 \pm 0.0050$  &  $0.4992 \pm 0.0069$  \\
     3.75  &  $0.5295 \pm 0.0053$  &  $0.5363 \pm 0.0053$  &  $0.4723 \pm 0.0068$  \\
     3.85  &  $0.5006 \pm 0.0062$  &  $0.5076 \pm 0.0062$  &  $0.4470 \pm 0.0072$  \\
     3.95  &  $0.4759 \pm 0.0061$  &  $0.4832 \pm 0.0062$  &  $0.4255 \pm 0.0071$  \\
     4.05  &  $0.4502 \pm 0.0064$  &  $0.4576 \pm 0.0065$  &  $0.4030 \pm 0.0071$  \\
     4.15  &  $0.4177 \pm 0.0069$  &  $0.4252 \pm 0.0070$  &  $0.3744 \pm 0.0074$  \\
     4.25  &  $0.4003 \pm 0.0071$  &  $0.4081 \pm 0.0073$  &  $0.3593 \pm 0.0075$  \\
     4.35  &  $0.3828 \pm 0.0105$  &  $0.3908 \pm 0.0107$  &  $0.3441 \pm 0.0102$  \\
     4.45  &  $0.3572 \pm 0.0097$  &  $0.3652 \pm 0.0099$  &  $0.3216 \pm 0.0094$  \\
     4.55  &  $0.3337 \pm 0.0110$  &  $0.3417 \pm 0.0112$  &  $0.3009 \pm 0.0104$  \\
     4.65  &  $0.3190 \pm 0.0125$  &  $0.3272 \pm 0.0129$  &  $0.2881 \pm 0.0117$  \\
     4.75  &  $0.2674 \pm 0.0219$  &  $0.2747 \pm 0.0225$  &  $0.2419 \pm 0.0201$  \\
     4.85  &  $0.2456 \pm 0.0164$  &  $0.2527 \pm 0.0168$  &  $0.2225 \pm 0.0151$  \\
   \hline
   \end{tabular*}
   \end{center}
\end{table}

\subsection{Analytic function}\label{sec:analytic}

\begin{figure*}
   \centering
   \begin{minipage}{\textwidth}
   \begin{center}
   \includegraphics[width=0.70\textwidth]{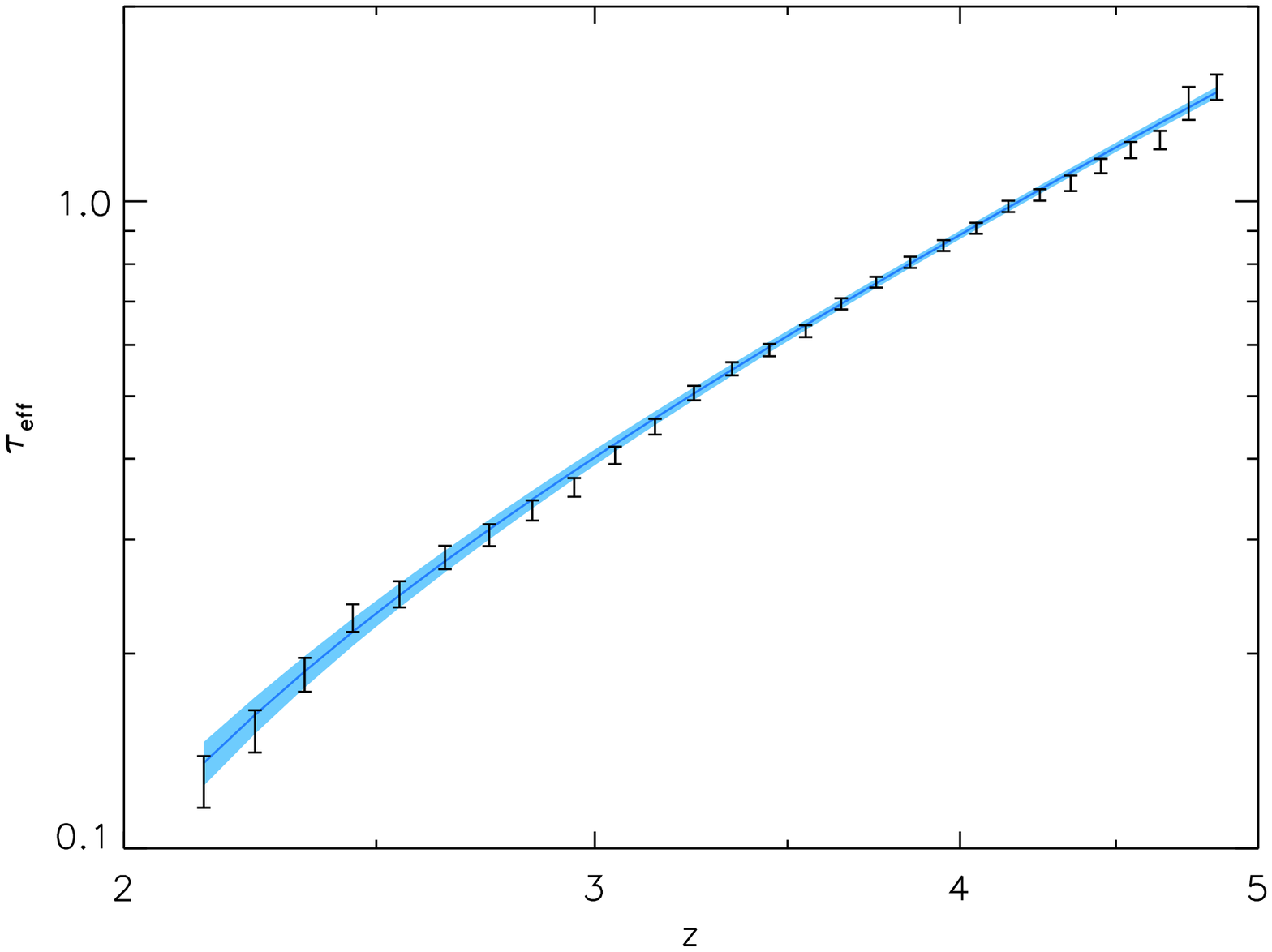}
   \vspace{-0.05in}
   \caption{Mean transmitted flux in the \lya\ forest expressed as an effective optical depth, $\tau_{\rm eff}(z) = -\ln{F(z)}$.  The solid line is an analytic fit to the full data set (not the binned points displayed), using a modified power law of the form $\tau_{\rm eff}(z) = \tau_0 \left[ (1+z)/(1+z_{0}) \right]^{\beta} + C$.  The shaded region gives the 1$\sigma$ uncertainty in the model \taueff\ values, taking into account the covariance between the model parameters.  See Section~\ref{sec:analytic} for details.  The axes are scaled such that a pure power law with $C = 0$ would appear as a straight line.}
   \label{fig:taueff}
   \end{center}
   \end{minipage}
\end{figure*}

In addition to fitting a discreet $F(z)$, we fit an analytic function to the effective optical depth as a function of redshift, $\tau_{\rm eff}(z) = -\ln{F(z)}$.  It has been conventional to parametrize \taueff\ as a simple power law of the form $\tau_{\rm eff}(z) = \tau_0 \left[(1+z)/(1+z_0)\right]^\beta$.  If one assumes this form for \taueffz, then both the normalization and the slope will be directly constrained from flux ratios, since neither $\tau_0$ nor $\beta$ cancels when computing $F(z_1)/F(z_2)$.  With the precision and long redshift baseline provided by our dataset, however, we found that, although a pure power law could provide a reasonable fit to the observed flux ratios (i.e., produce an $F(z)$ with the correct shape), the overall normalization of $F(z)$ was substantially too low when compared to literature values.  We found, however, that a reasonable fit could  be achieved by allowing for a constant offset to \taueffz, such that
\begin{equation}\label{eq:taueff}
\tau_{\rm eff}(z) = \tau_0 \left( \frac{1+z}{1+z_{0}} \right)^{\beta} + C \, .
\end{equation}
Following our approach for the discreet $F(z)$, we fit $\tau_0$ and $\beta$ directly using the flux ratios, then solved for $C$ by scaling our results to those of \citet{fg2008a} over $2.2 \le z \le 2.5$.  Bootstrapping was again used to compute error estimates, including the covariance between fit parameters.  

\begin{table}
   \caption{Covariance matrix for $\tau_{\rm eff}(z)$ parameters (equation~\ref{eq:taueff}).  The best-fitting values are $[\tau_0,\beta,C] = [0.751,2.90,-0.132]$, for $z_0 = 3.5$.\vspace{-0.1in}}
   \label{tab:taueff_covar}
   \begin{center}
   \begin{tabular*}{8.4cm} {@{\extracolsep{\fill}}cccc}
   \hline
      &  $\tau_0$  &  $\beta$  & $C$ \\
   \hline
      $\tau_0$  &   0.00049  &  -0.00241  &  -0.00043  \\
      $\beta$   &  -0.00241  &   0.01336  &   0.00224  \\
      $C$       &  -0.00043  &   0.00224  &   0.00049  \\
   \hline
   \end{tabular*}
   \end{center}
\end{table}

The results for \taueffz\ are plotted in Figure~\ref{fig:taueff}.  The best-fitting parameters are $[\tau_0,\beta,C] = [0.751,2.90,-0.132]$, for $z_0 = 3.5$, and the full covariance matrix is given in Table~\ref{tab:taueff_covar}.  We emphasize that $\tau_0$ and $\beta$ were constrained directly from the composite flux ratios (corrected for absorption due to optically thick systems) rather than the binned $F(z)$ values.  Nevertheless, the analytic fit closely matches the discreet values at all redshifts once both are scaled to the literature values at $z \sim 2$.    In $F(z)$ space, the deviations of the discreet $F(z)$ from the nominal analytic fit have a formal $\chi^2$ of 64, calculated using the full covariance matrix, for 28  redshift bins.  This suggests that some of the residual structure in the discreet fit may be real, that a modified power law is not a sufficiently accurate model for \taueffz, or that we have somewhat underestimated the errors in the discreet case.  Even so, the departures of the discreet points from the analytic fit are small, with a mean fractional deviation in the flux of two per cent.  A modified power law for \taueffz\ thus appears to be a reasonable approximation to the true evolution of the \lya\ opacity over this redshift range, although we caution that the fit should not necessarily be extrapolated to higher or lower redshifts.

\section{Comparison to Previous Results}\label{sec:comparison}

\begin{figure}
   \begin{center}
   \includegraphics[width=0.45\textwidth]{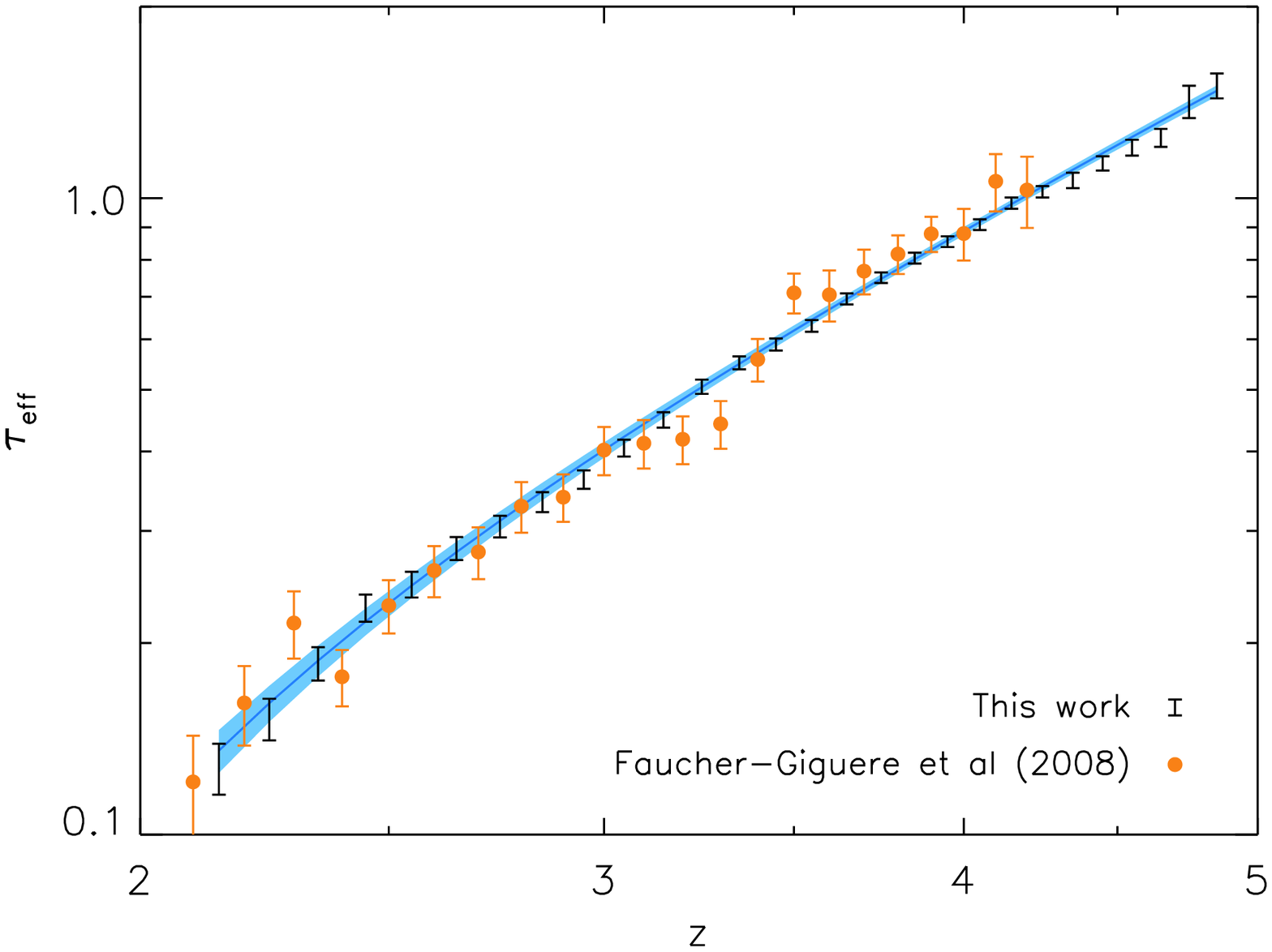}
   \vspace{-0.05in}
   \caption{Our results as shown in Figure~\ref{fig:taueff} plotted along with those of \citet{fg2008a}, which are the most careful measurements to date made with high-resolution spectra.  Although our values were scaled to the \fg\ measurements only at $z \le 2.5$, the overall agreement out to $z = 4.2$ is very good.  In light of the precision offered by the new measurements presented here, the feature at $z \sim 3.2$ tentatively identified by \fg\ appears to be consistent with noise.  We note that the \fg\ points over $2.2 \le z \le 2.5$, to which we normalized our results, appear noisy because they are presented on a log plot; their errors on the mean flux over this range are, in fact, relatively small ($\sigma_{F} \lesssim 0.02$).}
   \label{fig:taueff_with_fg08}
   \end{center}
\end{figure}

\begin{figure}
   \begin{center}
   \includegraphics[width=0.45\textwidth]{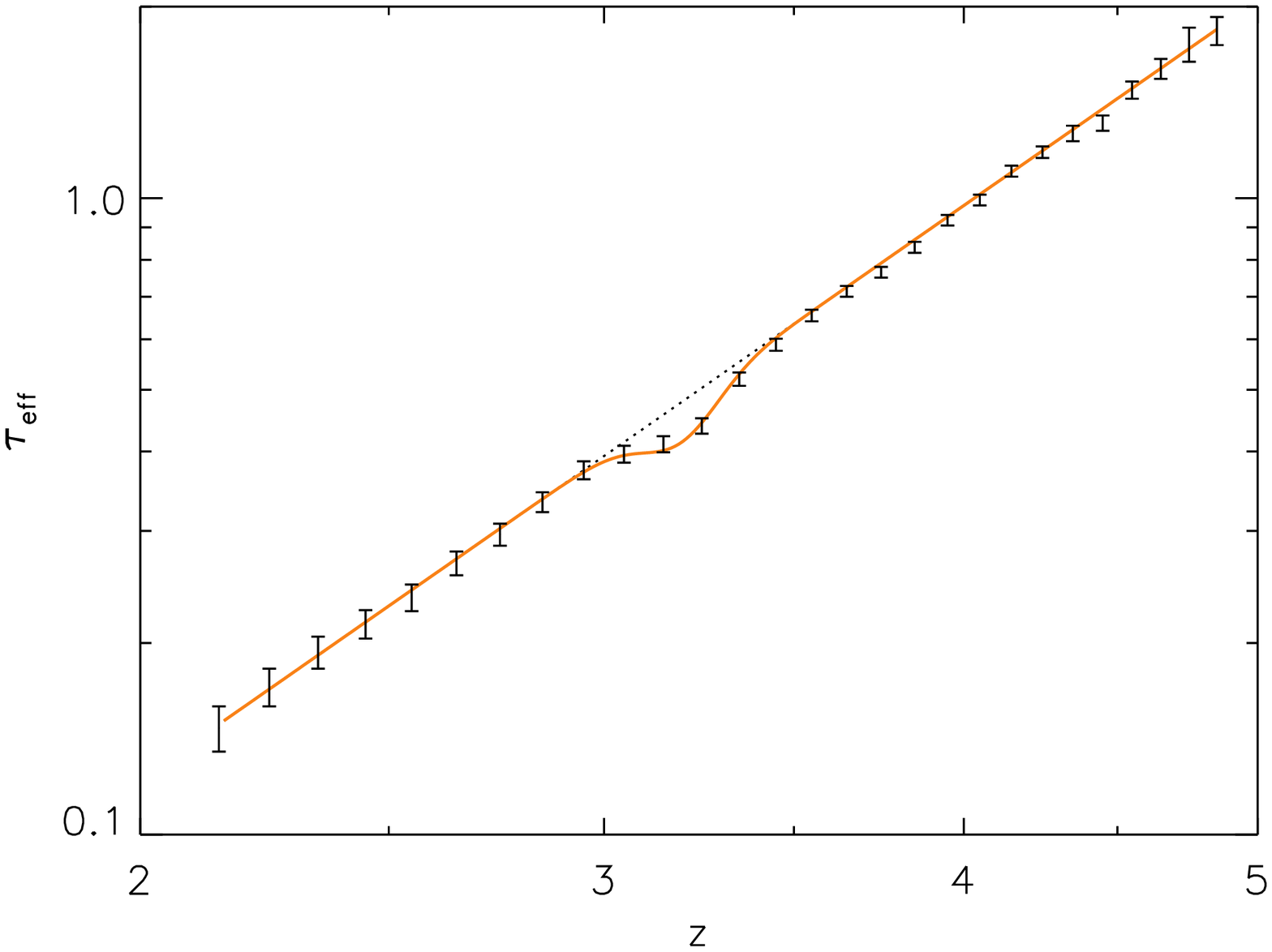}
   \vspace{-0.05in}
   \caption{A demonstration of the ability of our approach to recover a feature in the \taueffz\ evolution similar to the one reported by \citet{fg2008a} and \citet{bernardi2003}.  The solid line represents the input \taueffz, while the dotted line shows a pure power law for comparison.  Points with error bars are the \taueffz\ we recover by applying out analysis technique to artificial composite quasar spectra generated with the input \taueffz\ but with properties otherwise similar to those in our sample.  See Section~\ref{sec:comparison} for details.}
   \label{fig:taueff_recovery}
   \end{center}
\end{figure}

We now provide a brief comparison to some previous results from the literature.  The most careful measurements of the mean \lya\ flux using high-resolution spectra has arguably been made by \citet{fg2008a}, whose values at $z \le 2.5$ we have used to normalize our own results.  Their values are plotted along with our results in Figure~\ref{fig:taueff_with_fg08}.  Although we normalized our values to theirs only over $2.2 \le z \le 2.5$, the overall agreement out to $z = 4.2$ is very good.  We note that at $z \ge 3.5$ the \fg\ values for \taueffz\ are systematically somewhat higher ($\Delta \tau_{\rm eff} \sim 0.05$) than ours.  This may indicate that the continuum corrections made by those authors were somewhat too large at the high-redshift end.  More significantly, the feature in \taueffz\ at $z \simeq 3.2$ tentatively identified by \fg\ is absent from our results.  In light of the substantially higher precision offered by the present analysis, the feature appears likely to have been spurious.  It may either reflect random fluctuations in the \fg\ data, or may be the result of a subtle systematic effect.

A potential concern is whether an analysis using composite spectra, which average the flux from objects over a range of redshifts, may smooth out discreet features in the mean flux evolution.  We tested this by generating artificial composites with the same redshift range and binning as used for the real data, i.e., bins of $\Delta z = 0.1$ from $z = 2.2$ to 4.4, and $\Delta z = 0.4$ from $z=4.4$ to 5.2.  Each composite was generated from 100 individual artificial spectra with redshifts distributed uniformly across the bin.  For simplicity, a uniform continuum was assumed for each spectrum.  The input \lya\ forest absorption for each individual spectrum was calculated from the \citet{fg2008a} power law+Gaussian fit to \taueffz, which includes a Gaussian dip of width $\sigma_z \simeq 0.1$ centered at $z \simeq 3.2$\footnote{See values using the \citet{schaye2003} metal correction in Table 5 of \citet{fg2008a}.}.  This is comparable to the feature reported by \citet{bernardi2003}.  After adding noise to the artificial composites at a level comparable to that in the real data, we re-measured the mean flux evolution using the approach outlined in Section~\ref{sec:method}.  The results, plotted in Figure~\ref{fig:taueff_recovery}, show that the input flux evolution was recovered with high accuracy.  Although this was a simplified test, it demonstrates the key point that measurements of the mean flux evolution using composite spectra are entirely sensitive to discreet features similar to the ones that have been reported in the literature.  

\begin{figure*}
   \centering
   \begin{minipage}{\textwidth}
   \begin{center}
   \includegraphics[width=0.85\textwidth]{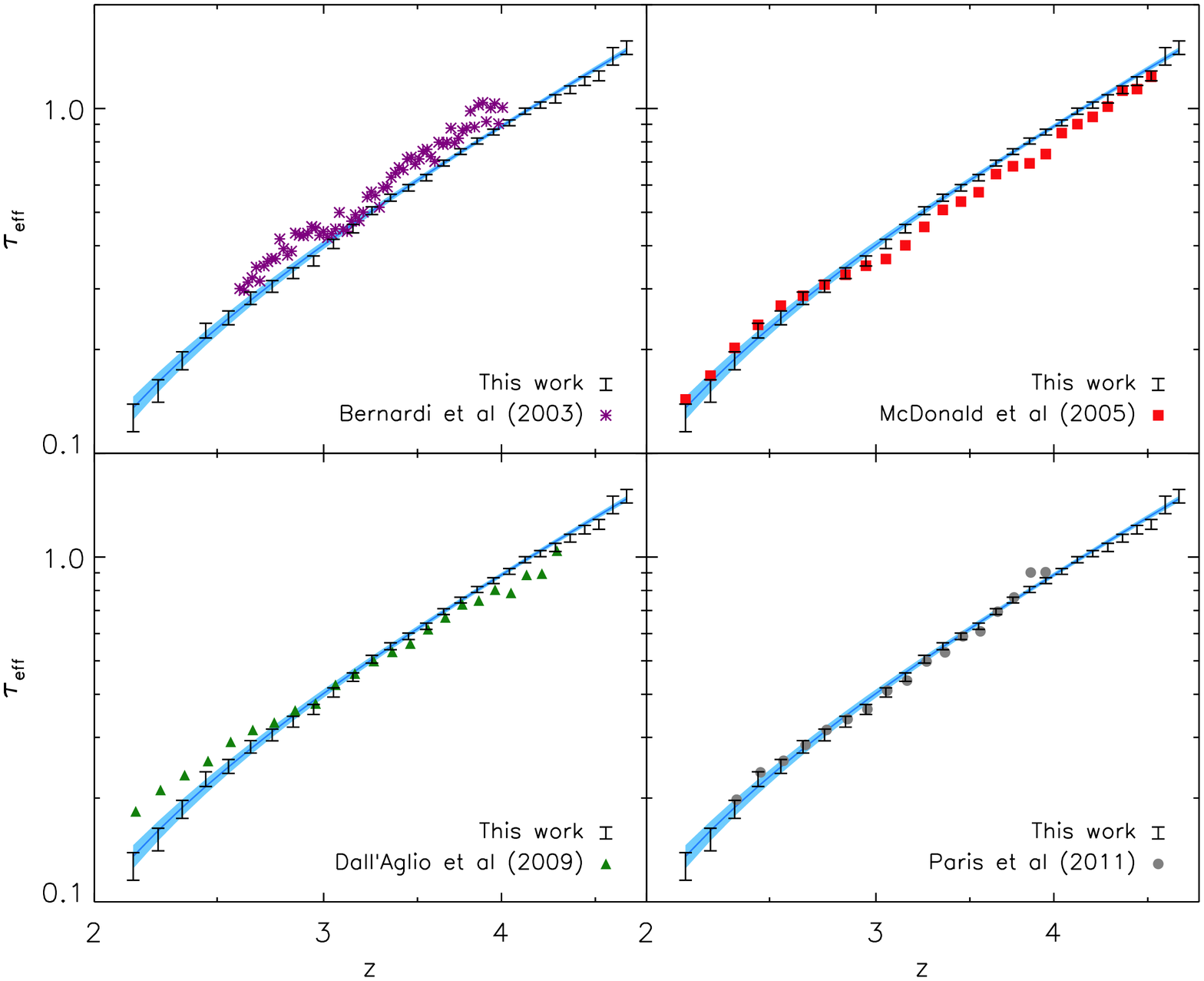}
   \vspace{-0.05in}
   \caption{Our results as shown in Figure~\ref{fig:taueff} plotted along with previous measurements made from SDSS spectra.  Error bars have been omitted from the literature values for clarity and to highlight systematic trends.  Deviations in the first three sets of measurements \citep{bernardi2003,mcdonald2005,dallaglio2009} from the results presented here are likely to reflect systematic errors in those works.  The apparent agreement between the \citet{paris2011} results and ours is generally good, although the \paris\ measurements have not been explicitly corrected for metal-line contamination.  See Section~\ref{sec:comparison} for discussion.}
   \label{fig:taueff_with_sdss}
   \end{center}
   \end{minipage}
\end{figure*}

We next turn to previous measurements made using SDSS spectra.  Results from \citet{bernardi2003}, \citet{mcdonald2005}, \citet{dallaglio2009}, and \citet{paris2011} are plotted along with our measurements in Figure~\ref{fig:taueff_with_sdss}.  In the three earliest works there are systematic differences from our results. We propose that these differences are most likely to be explained by systematic errors in the previous analyses, as discussed below.  In contrast, the apparent agreement between the \citet{paris2011} measurements and ours is generally very good.  At face value this suggests that recent results from SDSS spectra may have largely converged.  As we discuss below, however, some issues concerning the treatment of metals may remain with the \paris\ measurements.

\citet{bernardi2003} used composites generated from 1061 individual quasar spectra to simultaneously constrain \taueffz\ and the mean unabsorbed continuum over the \lya\ forest using two approaches.  In the first approach, they conducted a joint fit to analytic functions for both \taueffz\ and the mean continuum.  Second, they used an iterative approach to fit a discreet \taueffz\ along with an arbitrarily shaped continuum for each composite.  Both approaches, however, face significant limitations.  The composite spectra used by \bernardi, as well as the ones used here, fundamentally do not contain enough information to constrain both the amplitude and shape of $F(z)$ (or \taueffz) unless additional assumptions are made.  In their analytic approach, \bernardi\ assumed a power law + Gaussian form for \taueffz\ similar to the one used by \citet{fg2008a}.  As discussed in Section~\ref{sec:analytic}, the ratios of fluxes at different redshifts directly provided by composite spectra will constrain both the normalization and slope of a pure power law.  In this case, therefore, \taueffz\ will be implicitly normalized purely by the choice of fitting function (an additional, narrow Gaussian component will not change this since the contribution to \taueffz\ will go to zero far from the Gaussian centroid).  Based on the results presented here, we would expect that the \bernardi\ results would be off by at least a normalization factor, and indeed, an offset is seen in Figure~\ref{fig:taueff_with_sdss} that is consistent with the fact that we required a negative correction to \taueffz\ ($C < 0$) in order to correctly normalize our results.  We note that there may be additional systematic errors related to the choice of analytic shape for the continuum, which \bernardi\ parametrized as a power law plus three Gaussians to approximate the emission lines in the forest.  

For the second, iterative approach used by \citet{bernardi2003}, it is not clear that their results were converged, as claimed.   They used power laws normalized at 1450~\AA\ as initial estimates of the continua, and refined the continuum shapes based on the inferred $F(z)$ calculated at each iteration.  In tests with our data, we found that after 3--4 iterations, as used by \bernardi, this procedure roughly recovered the continuum shape plotted in Figure~\ref{fig:continuum}.  However, the normalization of the continuum, and hence $F(z)$, scaled directly with the normalization of the initial power-law continuum.  Moreover, although the changes in the continuum were minor after a few iterations, the continua were not converged.   After many iterations, the continua evolved to incorporate all of the relative differences between composites, producing a constant $F(z)$.  This appears to be a generic result when the continua are allowed to converge separately for each composite.  If the continua were held to be the same for each composite, then the shape of the continuum did converge, producing a relative $F(z)$ very similar to the one presented here.  The overall normalization, however, did not converge.  This simply demonstrates the fact that, without a completely unabsorbed composite to serve as a reference, the composites themselves do not contain enough information to fully constrain both $F(z)$ and the unabsorbed continuum without making other assumptions.

\citet{dallaglio2009} fitted continua to 1733 individual quasar spectra by combining localized spline fits with a redshift-dependent global correction estimated from mock spectra.  This approach is in principle similar to the one used by \citet{fg2008a}; however, there are at least two significant differences.  First, the use of lower-resolution data means that the corrections must be substantially larger, and will be significant even down to $z \sim 2$.  Additionally, the mock spectra used by \da\ were created by randomly generating absorption lines from fixed number density, column density, and Doppler parameter distributions, rather than produced from cosmological simulations, as used by \fg.  The continuum corrections estimated from these mock spectra will therefore be subject to inaccuracies in the line parameter distributions, which were themselves estimated from high-resolution spectra that may have suffered from continuum fitting errors.  \da\ also make the simplifying assumptions that the column density distribution is described by a single power law over $10^{12}~{\rm cm^{-2}} < N_{\rm H\, I} < 10^{18}~{\rm cm^{-2}}$ \citep[see][]{dallaglio2008}, and that the shapes of the column density and Doppler parameter distributions do not evolve with redshift.  Taken together, these systematic effects offer a reasonable explanation for the relative tilt in the evolution of \taueffz\ measured by \da\ with our respect to our values.  We note, however, that \da\ find a smooth evolution with redshift, similar to our results, without any evidence for the feature at $z \sim 3.2$ reported by \citet{bernardi2003}.  

\citet{mcdonald2005} also found a relatively smooth evolution of \taueffz\ with redshift in a sample of 3035 quasars.  We include this work for completeness, although those authors stress that their method for estimating individual quasar continua in the \lya\ forest using principle component analysis was not necessarily converged.  They also required a normalization factor taken from other measurements, though this would not affect the overall shape of \taueffz.  

Finally, \citet{paris2011} recently applied PCA to measure the mean transmitted \lya\ flux in the SDSS spectra of 2576 quasars.  Their approach used principle components derived from $z \sim 3$ quasar spectra to predict the unabsorbed continuum in the \lya\ forest of individual objects based on the flux redward of \lya.  Overall, their values are generally consistent with our results, although very mild systematic differences can be seen in Figure~\ref{fig:taueff_with_sdss}.  We caution that, despite the apparent agreement, a direct comparison with their results may not be entirely appropriate.  Although \paris\ addressed the issue of optically thick systems by  avoiding lines of sight containing DLAs, they did not explicitly correct for metal line contamination.  Since they were measuring the mean flux directly from continuum-normalized spectra, such a correction would have been necessary, in principle, to calculate the mean transmitted flux due to \lya\ alone.  We note that applying an adjustment of $\Delta \tau_{\rm eff} \simeq -0.03$ for metals (see Section~\ref{sec:metals}) would bring their \taueffz\ values systematically below our one-sigma limits everywhere except in their three highest redshift bins.  It is possible, however, that systematic effects in the continuum fitting  may have compensated for the lack of an explicit metal-line correction.  For example, when drawing the continuum over the \lya\ forest in their training set of quasar spectra, \paris\ generally followed the tops of the visible transmission peaks.  In moderate-resolution SDSS spectra, however, where absorption features in the forest are significantly smoothed, this procedure will tend to place the continuum slightly too low.  We note that the mean flux evolution over $2.5 \lesssim z \lesssim 2.7$ they measure from the training set is already within $\sim$1 per cent of the \citet{fg2008a} measurements, even though the \fg\ values were corrected for metals \citep[see][Figure 4]{paris2011}.  In addition, although visible metal lines redward of \lya\ were masked when fitting continua to the high-quality training spectra, the majority of spectra used for the mean flux measurement would have had significantly lower signal-to-noise ratios, making metals harder to identify.  This also may have caused the continua to be placed too low on average, potentially offsetting the absorption due to metals in the forest.  A detailed examination of the systematics involved in PCA measurements of the mean flux is beyond the scope of this work; however, effects such as those described above may potentially explain why the \paris\ results generally agree with our own, despite the fact that they do not perform an explicit correction for absorption due to metal lines.

\section{Summary}\label{sec:summary}

We have presented new measurements of the mean transmitted flux in the \lya\ forest over $2.15 \le z \le 4.85$ made using composite quasar spectra.  Our sample includes 6065 individual spectra from the SDSS DR7 quasar catalogue, which were combined into 26 composites with mean quasar redshifts $z_{\rm comp} = 2.25$ to 5.08.    

The fact that the quasar composites all have highly similar fluxes redward of \lya\ suggests that the differences in the observed flux in the forest are driven by the evolution in the mean intergalactic \lya\ opacity.  The ratio of $F(z)$ at two different redshifts can be measured by comparing the observed mean flux in two composites at the same rest wavelength.  We used these ratios to directly constrain $F(z)$ as a fraction of $F(z=2.15)$.  We also conducted tests that address our assumption that the composites have similar unabsorbed continua over the \lya\ forest, finding that the errors in $F(z)/F(z=2.15)$ arising from variations in the continua are likely to be within  our uncertainty estimates.  The overall normalization for $F(z)$ was determined using literature values made from high-resolution data at $z \le 2.5$ \citep{fg2008a}, where continuum uncertainties, although accounted for, are relatively minor.  The statistical accuracy offered by the large number of quasar spectra means that this normalization factor now dominates the uncertainty in $F(z)$ at $z \le 3.75$.  

Our main results give the mean transmitted flux associated with systems of column density $N_{\rm H\,I} < 10^{19}~{\rm cm^{-2}}$, and do not include the contribution from metal lines.  We fit the mean flux using two functions.  The first is a discreet $F(z)$ in bins of $\Delta z = 0.1$, while the second is a modified power law of the form $\tau_{\rm eff}(z) = \tau_0 \left[ (1+z)/(1+z_{0}) \right]^{\beta} + C$, where $\tau_{\rm eff}(z) = -\ln{F(z)}$.  Although previous works had generally parametrized \taueffz\ as a pure power law ($C=0$), we found that this was insufficient to describe the observed flux ratios at the required level of precision while simultaneously producing the correct normalization.  We emphasize that the true functional form of \taueffz\ is unknown, and that our parametrization is appropriate only in that it provides a reasonable match to the discreet $F(z)$ values.  Caution should be used when extrapolating this (or any) fit to higher and lower redshifts.  We also stress that, due to the fact that our results are generated from the ratios of fluxes at different redshifts, the parameters for both the discreet $F(z)$ and modified power law \taueffz\ are correlated, and that the full correlation matrices should be considered when applying our results.  

The mean transmitted flux is found to evolve smoothly with redshift, which is consistent with a gradual evolution of the \hi\ ionization rate and gas temperature over $2 < z < 5$.  We see no evidence of previously reported features at $z \simeq 3.2$, which appear to have been spurious.  Our results are otherwise generally consistent with the most careful measurements made to date using high-resolution data \citep{fg2008a}, but offer a substantially greater level of precision and extend out to higher redshifts.  Our measurements also offer an improvement over previous efforts using SDSS spectra, which appear generally to have been affected by systematic errors \citep[although see][]{paris2011}.

This work demonstrates the significant advantages of using large data sets to measure the mean transmitted flux.  In addition to the high degree of statistical accuracy, the use of composites makes it possible to avoid fitting continua at high redshifts, which is the primary source of systematic uncertainty in other approaches.  These measurements should therefore be useful to a variety of studies that use the \lya\ forest.

\section*{Acknowledgements} We thank Claude-Andr\'e Faucher-Gigu\`ere and Isabelle P\^aris for their helpful comments on a first draft of this paper.  We also thank the anonymous referee for helpful suggestions.  GDB gratefully acknowledges support from the Kavli Foundation.  PCH has received support from the STFC-funded Galaxy Formation and Evolution program at the Institute of Astronomy.  GW and JXP have been supported through an NSF CAREER grant (AST-0548180) and by NSF grant AST-0908910.

Funding for the SDSS and SDSS-II has been provided by the Alfred P. Sloan Foundation, the Participating Institutions, the National Science Foundation, the U.S. Department of Energy, the National Aeronautics and Space Administration, the Japanese Monbukagakusho, the Max Planck Society, and the Higher Education Funding Council for England. The SDSS Web Site is http://www.sdss.org/.

The SDSS is managed by the Astrophysical Research Consortium for the Participating Institutions. The Participating Institutions are the American Museum of Natural History, Astrophysical Institute Potsdam, University of Basel, University of Cambridge, Case Western Reserve University, University of Chicago, Drexel University, Fermilab, the Institute for Advanced Study, the Japan Participation Group, Johns Hopkins University, the Joint Institute for Nuclear Astrophysics, the Kavli Institute for Particle Astrophysics and Cosmology, the Korean Scientist Group, the Chinese Academy of Sciences (LAMOST), Los Alamos National Laboratory, the Max-Planck-Institute for Astronomy (MPIA), the Max-Planck-Institute for Astrophysics (MPA), New Mexico State University, Ohio State University, University of Pittsburgh, University of Portsmouth, Princeton University, the United States Naval Observatory, and the University of Washington.

\bibliographystyle{apj}
\bibliography{mean_flux_refs}

\end{document}